\documentclass[11pt]{article}
\usepackage[margin=0.95in]{geometry}
\usepackage{epsfig,ulem,amssymb,soul}

\usepackage{enumitem} 
\usepackage{siunitx}
\usepackage{epstopdf}
\usepackage{amsfonts,amsmath,braket}
\usepackage{bm}
\usepackage{setspace}
\usepackage{comment}
\usepackage{siunitx}
\usepackage{mathrsfs}
\usepackage{multirow}
\usepackage{subcaption}
\usepackage[hang,flushmargin,bottom]{footmisc}
\usepackage{mathtools}
\usepackage{lipsum}
\usepackage{array}
\usepackage{ifthen}
\usepackage[affil-it]{authblk}
\usepackage{cite}
\usepackage[table]{xcolor}
\usepackage{leftidx}
\usepackage{relsize}
\usepackage{braket}
\usepackage{float}
\usepackage{placeins}
\usepackage[colorlinks=true, linkcolor=blue, citecolor=blue]{hyperref} 

\usepackage[autostyle]{csquotes}
\usepackage[dvipsnames]{xcolor}

\usepackage{hyperref}

\title{A Spectral Quantum Algorithm for Numerical Differentiation and Integration}
\date{}

\author{Jordan Cioni, Fabio Semperlotti\thanks{To whom correspondence should be addressed. Email: fsemperl@purdue.edu }}
\affil{Ray W. Herrick Laboratories, School of Mechanical Engineering,  Purdue University, West Lafayette, IN 47907, USA}
\setstcolor{red}

\usepackage{lineno}

\begin{document}
\maketitle

\begin{abstract}
\noindent
Numerical calculus algorithms, which estimate derivatives and integrals from data series acquired either via measurements or by sampling functions, are essential in scientific computing.
To date, a few quantum algorithms have been developed to perform calculus operations based on closed-form functional inputs; yet, in many practical applications, field variables are numerically described via a series of samples rather than closed-form expressions.
This paper presents the theoretical development and the gate-level circuit implementation of quantum algorithms for numerical differentiation and, up to addition of a constant, indefinite integration.
The methodology relies on a spectral approach that leverages the computational efficiency of the quantum Fourier transform and the parallel computing capability afforded by quantum superposition 
to evaluate outputs at all domain points simultaneously.
The differentiation approach is also extended to enable gradient estimation, and post-processing procedures are presented to recover sign information. 
The primary output of the proposed algorithms are quantum state vectors directly proportional to the numerical derivative or integral of the given data; therefore, the correctly signed results are made available to proceeding quantum computations.
This result lays the foundation for the proposed algorithms to serve as core subroutines in applied quantum computing operations such as image processing, data analysis, and machine learning.

\end{abstract}

\section{Introduction}

As the field of fault-tolerant quantum computing rapidly expands, there is an increasing demand for quantum computing algorithms with practical applications.
This work focuses on quantum calculus algorithms, which carry significant practical value across the spectrum of scientific computing.
Many problem-specific quantum calculus algorithms have been developed within the disciplines of
quantum chemistry \cite{hohenstein_hybrid_2019, thomas_e_obrien_calculating_2019, mitarai_theory_2020}, optimization \cite{chan_proceedings_2019, chen_pure_2023}, finance \cite{colucci_quantum_2021, herbert_quantum_2022}, and image processing \cite{johnston_quantum_2016, shimada_quantum_2020}.
More specifically, in the context of quantum algorithms for function differentiation, three techniques are commonly employed: Algorithmic Differentiation (AD) \cite{colucci_quantum_2021}, Variational Quantum Eigensolvers (VQE) \cite{hohenstein_hybrid_2019, thomas_e_obrien_calculating_2019, mitarai_theory_2020}, and Jordan’s algorithm \cite{jordan_fast_2005, chan_proceedings_2019}.
In the context of quantum integration, research efforts have focused primarily on implementation of Quantum Monte Carlo Integration (QMCI) methods \cite{herbert_quantum_2022, shimada_quantum_2020, johnston_quantum_2016}; however, Riemann sum methods have recently been explored \cite{shukla_efficient_2024}.

In this section, we briefly review the current state of the art of quantum algorithms for differentiation and integration and we identify key shortcomings.
In the remaining sections of this paper, we propose a methodology that addresses these gaps and provides algorithms for quantum differentiation and integration that require minimal a-priori knowledge, operate directly using data samples, and return domain-wide differentiation and integration results simultaneously.

\subsection{Background on Quantum Differentiation Algorithms}
Algorithmic Differentiation (AD) is a computational implementation of the chain rule \cite{colucci_quantum_2021} and, as such, it is not accompanied by the approximation errors of traditional numerical methods \cite{bhaskar_quantum_2016}.
In both quantum and classical implementations, AD takes an input function and a point within its domain, and reduces it to a composition of primitive functions and operations (i.e. logarithms, trigonometric functions, addition, multiplication, etc.) with known derivatives \cite{neidinger_introduction_2010, henrard_algorithmic_2017}.
Using quantum implementations of the primitive functions \cite{cao_quantum_2013, bhaskar_quantum_2016, hadfield_quantum_2018}, the derivative is evaluated at the given point \cite{colucci_quantum_2021}.
It follows that the error associated with this technique is primarily associated with the quantum implementation of each primitive function \cite{bhaskar_quantum_2016}.
Likewise, its complexity is given by the sum of each primitive function’s quantum complexity \cite{colucci_quantum_2021}.
Compared to other numerical differentiation techniques, AD is not capable of estimating a derivative from a set of function samples taken at points in the function domain. 
Furthermore, because AD only returns the input function’s derivative at a single point, the technique must be repeatedly applied to estimate a function’s derivative throughout a computational domain.

In the field of quantum chemistry, the VQE is commonly used to estimate energy derivatives with respect to atomic variables such as nuclei position and external fields \cite{mitarai_theory_2020, schlegel_geometry_2011, wilson_molecular_1955, thomas_e_obrien_calculating_2019}, which correspond to a variety of physical properties including nuclear forces, vibrational frequencies, and radiation intensities \cite{schlegel_geometry_2011, wilson_molecular_1955}. 
VQEs use a parameterized circuit, $U(\theta)$, and a corresponding ansatz state, $\ket{\psi(\theta)} = U(\theta)\ket{\psi_0}^{\bigotimes n}$, where $\ket{\psi_0}^{\bigotimes n}$ is an $n$-qubit initial state vector, and $\theta$ is a vector of circuit parameters.
The parameters are variationally optimized until the expectation value of the circuit, $E$, is minimized. 
Then, the derivative of the expectation value is obtained from previously known relationships with the optimal circuit parameters
\cite{mitarai_theory_2020,pulay_analytical_1987}.
For this approach to be applicable, the differentiand must be encoded as the expectation value of a Hamiltonian operation. For this reason, the approach is especially suitable for studying atomic energy derivatives because the Hamiltonian operator $\hat{H} = i \hbar \frac{\partial}{\partial x}+V(x)$, where $V(x)$ is the potential energy of the system, already corresponds to an atomic system's total energy.
Given that this quantum derivative technique requires problem-specific information to encode a suitable Hamiltonian operator, it is not readily generalized to other problems and disciplines. 
Due to the analytical nature of this technique \cite{mitarai_theory_2020}, the VQE approach requires prior knowledge of the input’s functional form; therefore, this method also cannot estimate a derivative from a set of samples. 

Jordan’s algorithm for numerical gradient estimation\cite{jordan_fast_2005} approximates the input function, $f$, over a small domain centered at the origin using a first-order Taylor series, $f(\vec{x}) = f(\vec{0})+\vec{x}\cdot\nabla f$.
The input function is evaluated using an oracular function, $O_f = e^{i2\pi\frac{N}{ml}f(\vec{x})}$, where $l$ defines a sufficiently small region to enable a linear approximation of $f$, $m$ is an estimate of the largest partial derivative, and $N=2^n$ where $n$ is the number of qubits used to describe the partial derivatives along each dimension.
For a $d$-dimensional function, a single query is made to the oracle function to evaluate $f$ at $d$ points, one along each dimension, surrounding the origin.
Using addition modular $N$, the result is added to an output register prepared in an eigenstate of addition mod $N$.
Consequently, the input registers receive a phase-kickback proportional to $f$\cite{wong_introduction_2022}.
Recalling the first-order Taylor expansion of $f$, the global phase of $f(\vec{0})$ is ignored, and the phase terms proportional the function's gradient are extracted using a QFT. 

Using the implementation of Jordan’s algorithm described above, a function’s gradient may be estimated efficiently; however, it is limited to returning gradient estimations at the origin of the input domain \cite{jordan_fast_2005}.
Owing to the prevalence of gradient descent methods in variational quantum algorithms and Quantum Approximate Optimization Algorithms (QAOA) \cite{mcclean_theory_2016, wecker_progress_2015}, efforts have been made to expand the capabilities of Jordan’s algorithm for these purposes \cite{chen_pure_2023, chan_proceedings_2019}.
Using separable variable quantum addition and subtraction techniques, the input state is shifted by an arbitrary $\Delta x$, allowing Jordan’s algorithm to be used for a region centered beyond the origin of the input domain \cite{chen_pure_2023}.
Consequently, fully quantum gradient descent is achievable \cite{chen_pure_2023}.
Noting that Jordan’s algorithm encodes the objective function using a single call to a phase oracle function \cite{jordan_fast_2005}, the gate-level complexity of the algorithm is undetermined without an estimate of the phase oracle's complexity.
Efforts to derive an efficient phase oracle for Jordan’s algorithm have shown that, to evaluate a general $d$-dimensional function, $f$, with precision $\epsilon$, an oracle may be constructed with complexity $\mathcal{O}\big(\frac{\sqrt{d}}{\epsilon ^2}\big)$ \cite{chan_proceedings_2019}; however, for smooth functions, the same oracle has complexity $\mathcal{O}\big(\frac{\sqrt{d}}{\epsilon}\big)$ \cite{chan_proceedings_2019, gilyen_quantum_2019}.
Recalling that Jordan’s algorithm relies on first-order terms of the given function’s Taylor expansion \cite{jordan_fast_2005}, it has been shown that incorporation of higher-degree central difference formulas enables extension of Jordan’s algorithm to functions possessing non-negligible higher-order terms \cite{chan_proceedings_2019}. 

Based on the above discussion of currently available algorithms for quantum differentiation, it is found that most existing methods require significant prior knowledge of the input function. 
AD requires the functional form of the input to be known \cite{colucci_quantum_2021}, VQE approaches require a known input function encoded as a Hamiltonian \cite{mitarai_theory_2020}, and Jordan’s algorithm requires a phase oracle constructed from the input function \cite{jordan_fast_2005}. 
At the same time, no existing techniques exploit quantum entanglement to return simultaneous derivative estimates at multiple points.
It appears that quantum algorithms capable of performing numerical derivatives in a traditional sense, that is taking function samples as input and returning the derivative estimates as output without additional prior knowledge, are needed.

\subsection{Background on Quantum Integration Algorithms}

A classical Monte Carlo Integration (MCI) approximates the values of bound integrals by estimating the function’s mean over a given domain using successive function evaluations at random points.
The integral value, which is proportional to the function’s mean, is easily obtained \cite{sugiyama_numerical_2016}.
In quantum implementations, QMCI follows a nearly identical framework.
Several researchers have employed quantum computing to improve upon the $\mathcal{O}\big(N\big)$ complexity of classical MCI, where integration is performed over $N$ discrete domain points \cite{abrams_daniel_s_fast_1999, johnston_quantum_2016, shimada_quantum_2020, acioli_review_1997, heinrich_quantum_2003, shu_general_2024}.
Using Grover’s algorithm to improve mean estimation resulted in a quadratically improved complexity of $\mathcal{O}\big(\sqrt{N}\big)$; however, this implementation requires a Boolean input function \cite{abrams_daniel_s_fast_1999} and efficiency of the Grover’s algorithm step critically depends on the function’s uniformity over the input domain \cite{sun_quantum_2024}. 
In the Quantum Supersampling method, QMCI successfully improved efficiency by simultaneously evaluating the input function at all discretized domain points stored in a superposition of states \cite{johnston_quantum_2016}.
Building on the Quantum Supersampling method, the quantum coin method was developed for functions that map inputs to $\{0,1\}$ with quadratic improvements on the classical MCI complexity\cite{shimada_quantum_2020}.
Similar complexity improvements are also seen in QMCI algorithms developed for periodic functions \cite{acioli_review_1997}, Sobolev-like higher-dimensional functions \cite{heinrich_quantum_2003}, and Holder class functions \cite{dewitt-morette_path_1979}.
Noting that all of these examples strictly limit the input function, a generalized QMCI algorithm with similar $\mathcal{O}\big(\sqrt{N}\big)$ complexity was recently developed using oracular polynomial function approximations to simultaneously evaluate the function over a superposition of discrete domain points \cite{shu_general_2024}.
As highlighted for the quantum derivative algorithms, QMCI algorithms require significant prior knowledge of the problem. In particular, the integrand must be known in advance. These integration algorithms also restrict the type of integrations that may be performed: namely, they only perform definite numerical integration between specified limits.

Recently, a new quantum integration algorithm based on Riemann summations has been suggested as an alternative to QMCI \cite{shukla_efficient_2024}.
This novel quantum integration technique assumes normalized samples of the input function, $f$, at uniformly spaced domain points, $x_j$, can be efficiently encoded into the $N$ state amplitudes of an $n$-qubit system, where $N=2^n$.
The integral over the first $M$ terms, where $1 \leq M \leq N$, may be approximated by a Riemann sum, $\Delta x \sum_0^{M-1} f(x_j)$.
In this quantum algorithm, the partial summation is performed using a single unitary operation with complexity $\mathcal{O}\big(\log M\big)$ for a general value of $M$.
The unitary operator's circuit is designed such that the corresponding matrix's first column contains only ones in the first $M$ rows while the remaining rows contain only zero.
The application of the operator stores the result of the partial summation in the amplitude of the system's first state.
Once the first state is measured, the result is scaled by $\Delta x$ and the input function's $l^2$ norm during post-processing to recover the correct integral estimate.
Compared to the $\mathcal{O}\big(N\big)$ complexity of common classical numerical integration algorithms, and the $\mathcal{O}\big(\sqrt{N}\big)$ \cite{shu_general_2024} complexity of the most efficient QMCI algorithms, the quantum Riemann sum algorithm's maximum complexity of $\mathcal{O}\big(\log N\big)$ \cite{shukla_efficient_2024} represents a profound efficiency improvement which enables this algorithm to serve as a subroutine of more complex quantum algorithms. The primary limitation of this algorithm stems from its inability to estimate all partial summations simultaneously. In cases where the integral is required at all points along the domain, rather than a single exact integral estimate between two specified limits, this algorithm must be applied iteratively and will incur a complexity of $\mathcal{O}\big(M\log M\big)$.

Based on the above considerations, it appears that there is only one algorithm capable of estimating the integral from a set of function samples, while no methods are available to evaluate the integral point-wise over the computational domain.
In this work, we present a technique for computing partially bound integrals, which yield the integral value at all domain discretizations between the specified limits.
Additionally, this work's contributions to quantum integral calculation build off of its quantum derivative results to demonstrate a quantum implementation of the trapezoidal numerical integration technique, potentially demonstrating a route to developing higher-order quadrature-based quantum integration algorithms.

\section{Overview of the Technical Approach}

The quantum derivative algorithm presented in this paper builds on the properties of the Fourier transform and re-envisions a classical algorithm originally presented in \cite{johnson_notes_nodate, sunaina_calculating_2018}. Given a complex-valued integrable Lebesgue function $f: R \to \mathbb{C}$ defined on the real line such that $f \equiv f(x)$, the Fourier transform $\mathcal{F}$ of the derivative of $f(x)$ is given by \cite{kreyszig_advanced_2011}:
\begin{equation}\label{eq: iOmegaF}
    \mathcal{F}\bigg[\frac{df}{dx}\bigg]=i \omega \mathcal{F}\big[f(x)\big]
\end{equation}
where $i = \sqrt{-1}$ denotes the imaginary unit, and $\omega$ takes the meaning of either the pulsation or the wavenumber depending on the nature of the independent variable (either time or space, respectively). 
By applying the inverse Fourier transform to Eq.~\eqref{eq: iOmegaF}, the function's derivative is computed purely via multiplication and Fourier transform operations:
\begin{equation}\label{eq: FiOmegaF}
    \frac{df}{dx}=\mathcal{F}^{-1}\bigg[i \omega \mathcal{F}[f(x)]\bigg]
\end{equation}

While Eq.~\eqref{eq: FiOmegaF} represents a trivial technique for the calculation of analytical derivatives and a computationally expensive method for the classical calculation of numerical derivatives, it provides an effective blueprint for a quantum derivative algorithm.
In fact, the methodology described above can form the foundation of an efficient quantum derivative algorithm, given that the Quantum Fourier Transform (QFT) is, to-date, one of the most efficient quantum algorithms available.  

In the following, we will describe how Eq.~\eqref{eq: FiOmegaF} may be applied in discrete contexts for numerical differentiation and integration. Then, we will provide the mathematical description of both the Quantum Fourier Transform-based Derivative (QFTD) and the Quantum Fourier Transform-based Integral (QFTI) algorithms. 
We highlight that, with appropriate encoding and bit ordering, the derivative algorithm presented herein may be extended to higher-dimensional problems. 
Consequently, we briefly present the methodology for higher-dimensional QFTD operations.
Both the QFTD and QFTI successfully encode derivative or integral results in quantum state amplitudes; however, sign information is lost during the measuring process.
We end this section with a discussion of techniques that may be employed to recover sign information from QFTD and QFTI results.
For each algorithm, it will be assumed that the input function $f$ can be efficiently encoded using amplitude encoding, or that $f$ is encoded as a result of previous circuit operations.
In all derivations, we employ little endian convention and all circuit depictions include the least significant qubit in the top-most position.

\subsection{Discrete Fourier Transform-based Derivatives}
By leveraging the Discrete Fourier Transform (DFT), researchers have implemented classical derivative algorithms based on the discrete analog of Eq.~\eqref{eq: FiOmegaF} \cite{johnson_notes_nodate, sunaina_calculating_2018}. In this latter case, Eq.~\eqref{eq: FiOmegaF} is approximated as:
\begin{equation}\label{eq: DFT_FiOmegaF0}
    \bigg[\frac{\Delta f}{\Delta x}\bigg]_j = DFT^{-1}\big[i\omega_k ~DFT[f]_k\big]_j,
        \qquad \text{where }\omega_k = 
    \begin{cases}
        2\pi k/N\Delta x, & \quad 0\leq k< \frac{N}{2}\\
        2 \pi (k-N)/N\Delta x, & \quad \frac{N}{2} \leq k < N        
    \end{cases}
    \,.
\end{equation}
where $\big[ \frac{\Delta f}{\Delta x} \big]_j$ denotes the approximate derivative at the point $x_j$, given by $x_j=j\Delta x$, $\omega_k$ denotes the $k^{\text{th}}$ discrete equivalent of the wavenumber from Eq.~\eqref{eq: iOmegaF}, and the discrete Fourier transform and its inverse are defined as follows:
\begin{equation}\label{eq: DftDefn}
\begin{split}
    DFT[f]_k &=
    \frac{1}{\sqrt{N}} \sum_{j=0}^{N-1}f(x_j)\exp(-i\omega_k x_j)\\
    DFT^{-1}[F]_j &=
    \frac{1}{\sqrt{N}} \sum_{k=0}^{N-1}F(\omega_k)\exp(i\omega_k x_j)
\end{split}
\end{equation}

Noting that the identity provided in Eq.~\eqref{eq: iOmegaF} is obtained by analytically evaluating the Fourier transform of a function's derivative, the identity cannot be immediately extended to discrete applications where the function is known only numerically. Consequently, erroneous high-frequency oscillations were observed near domain boundaries when Eq.~\eqref{eq: DFT_FiOmegaF0} was applied to calculate numerical derivatives \cite{johnson_notes_nodate, sunaina_calculating_2018}.
In the case of non-periodic input functions, such Gibbs phenomena was exacerbated.
These works also found that end behavior discrepancies are minimized by adapting Eq.~\eqref{eq: DFT_FiOmegaF0} to use a modified wavenumber, denoted $\tilde{\omega}$. While this modified wavenumber is well-documented in the existing literature \cite{johnson_notes_nodate, sunaina_calculating_2018}, its derivation exemplifies a robust process for development of QFT-based quantum algorithms. As such, a brief summary of its derivation will be presented.

To obtain the modified wavenumber appropriate for discrete Fourier transform-based derivatives, we begin with the central difference method. 
For a function $f(x)$ sampled at discrete points $x_j$ with uniform separation given by $\Delta x$, the central difference method approximates the derivative at each point $x$ with the expression:
\begin{equation}\label{eq: CentralDiff}
    \bigg[\frac{\Delta f}{\Delta x}\bigg] = \frac{f(x + \Delta x) - f(x - \Delta x)}{2 \Delta x}
\end{equation}

Taking the Fourier transform of Eq.~\eqref{eq: CentralDiff} and applying the Fourier shift theorem, which states $\mathcal{F}\big[f(x-a)\big]_k=e^{-i\omega_k a}\mathcal{F}\big[f(x)\big]_k$, we obtain the relationship:
\begin{equation}\label{eq: ModW_Fk}
    \begin{split}
    \mathcal{F} \bigg[ \frac{\Delta f}{\Delta x} \bigg]_k & = 
    \frac{e^{i \omega_k \Delta x} - e^{-i \omega_k \Delta x}}{2\Delta x}\mathcal{F}_k 
    = \frac{i\sin(\omega_k \Delta x)}{\Delta x} \mathcal{F}_k
    \end{split}
\end{equation}
where $\mathcal{F}\big[f(x)\big]_k$ is abbreviated as $\mathcal{F}_k$ and the definition of $\omega_k$ is given in Eq.~\eqref{eq: DFT_FiOmegaF0}.
We now wish to rewrite Eq.~\eqref{eq: ModW_Fk} by substituting the definition of $\omega_k$ from Eq.~\eqref{eq: DFT_FiOmegaF0}.
Noting that $\sin\big(\frac{2\pi (k-N)}{N\Delta x}\big) = \sin\big(\frac{2\pi k}{N\Delta x}\big)$, Eq.~\eqref{eq: ModW_Fk} does not need to be treated in a piecewise manner to account for the piecewise nature of $\omega_k$.
Without loss of generality, $\omega_k$ is taken to be $\frac{2\pi k}{N\Delta x}$ when evaluating Eq.~\eqref{eq: ModW_Fk} and the modified wavenumber, $\tilde{\omega}_k$, simplifies to $\frac{1}{\Delta x}\sin(\frac{2\pi k}{N})$.
Consequently, the discrete equivalent of the Eq.~\eqref{eq: FiOmegaF} is obtained:
\begin{equation}\label{eq: Discrete_FiOmegaF}
    \bigg[\frac{\Delta f}{\Delta x}\bigg]_j =
    DFT^{-1}\big[
    i\tilde{\omega}_k
    DFT[f]_k
    \big]_j,
    \qquad \text{where }\tilde{\omega}_k =\frac{\sin(\frac{2\pi k}{N})}{\Delta x}
\end{equation}

\subsection{DFT-based Integration}
Using a procedure similar to that shown above for DFT-based differentiation, a modified wavenumber may also be obtained for DFT-based integration.
Beginning with the discrete approximation of the area beneath a function $f$ given by the trapezoidal method:
\begin{equation}\label{eq: TrapMeth}
    \Delta A_j = \frac{\Delta x}{2}\big[f(x_j+\Delta x)+f(x_j-\Delta x)\big]
\end{equation}
where $\Delta A_j$ is the area beneath the $j^\text{th}$ section of $f$.
As before, applying a Fourier transform to Eq.~\eqref{eq: TrapMeth} results in a trigonometric relationship:
\begin{equation}
    \mathcal{F}[\Delta A]_k
    = 
    \frac{\Delta x}{2}\big[e^{i\omega_k \Delta x} + e^{-i\omega_k \Delta x}\big]\mathcal{F}_k 
    =
    \Delta x\cos\bigg(\frac{2\pi k}{N}\bigg)\mathcal{F}_k
\end{equation}
Denoting the modified wavenumber for DFT-based integration by $\tilde{\omega}'$, the differential area under the $j^{th}$ section of $f$ may be represented as:
\begin{equation}\label{eq: DiffArea}
    \Delta A_j = DFT^{-1}\big[
    \tilde{\omega}'_k DFT[f]_k
    \big]_j,
    \qquad \text{where }\tilde{\omega}'_k=\Delta x\cos\bigg(\frac{2\pi k}{N}\bigg)
\end{equation}

Up to this point, the DFT-based derivative and integral procedures are nearly identical: both operations convert the input function to the frequency domain, scale by a modified wavenumber, and convert back to original domain.
While this process was sufficient for the derivative, the integral procedure requires an additional step.
In the continuous case, an integral with a fixed lower bound and an upper bound located arbitrarily within the problem domain is given by:
\begin{equation}\label{eq: ContinuousIntegral}
    \begin{split}
    \int_{x_0}^x f(x)~dx &= F(x) - F(x_0)\\
    &= I(x)
    \end{split}
\end{equation}
where $F(x)$ is the functional form of the integral of $f(x)$, $F(x_0)$ is the constant of integration obtained by evaluating $F(x)$ at $x_0$, and $I(x)$ is introduced to denote the overall integrated result $F(x)-F(x_0)$.
This expression is equivalent to an indefinite integral of the form $\int f(x) dx=F(x) + C$, where the constant of integration is chosen equal to $-F(x_0)$.
Evaluating Eq.~\eqref{eq: ContinuousIntegral} for a particular value of $x$ yields the area under $f$ between $x$ and $x_0$. 
Therefore, to discretely approximate the result of Eq.~\eqref{eq: ContinuousIntegral} at an arbitrary $x_j$ in the problem domain, a partial summation is performed from $x_0$ to $x_j$:
\begin{equation}
    I(x_j) = \sum_{i=0}^j\Delta A_i
\end{equation}
Finally, using a unit lower triangular matrix denoted by $\Sigma$, a single matrix product operation may be applied to evaluate $I(x_j)$, abbreviated $I_j$, for all domain values:
\begin{equation}\label{eq: LowerTri_mat}
    \begin{split}
        \begin{pmatrix}
            I_0 \\ I_1 \\ \vdots \\ I_{N - 1} 
        \end{pmatrix}
        & = \Sigma \vec{A}
        =
        \begin{bmatrix}
            1 & 0 & \cdots & 0 \\
            1 & 1 & \cdots & 0 \\
            \vdots & \vdots & \ddots & \vdots \\
            1 & 1 & \cdots & 1
        \end{bmatrix}
        \begin{pmatrix}
            \Delta A_0 \\ \Delta A_2 \\ \vdots \\ \Delta A_{N - 1}
        \end{pmatrix} \\
    \end{split}
\end{equation}

Combining Eq.~\eqref{eq: LowerTri_mat} and Eq.~\eqref{eq: DiffArea}, we obtain a discrete spectral method for approximating Eq.~\eqref{eq: ContinuousIntegral}, which defines an indefinite integral with a prescribed integration constant of $C=-F(x_0)$:
\begin{equation}
    I_j = \Sigma ~DFT^{-1}\big[
    \tilde{\omega}'_k DFT[f]_k
    \big]_j
\end{equation}

\subsection{The Quantum Fourier Transform Derivative (QFTD)}

In this section, we present the theory and corresponding quantum circuit implementation of the QFTD algorithm.
The basic QFTD algorithm includes three key steps: 1) mapping the input function to the frequency space, 2) scaling it by an appropriate modified wavenumber, and 3) mapping the result back to the input domain.
In the context of quantum computing, where the QFT is used instead of the DFT, the modified wavenumber used to calculate a derivative is identical to the classical case.
\begin{figure}[!h]
    \centering
    \includegraphics[width=0.8\linewidth]{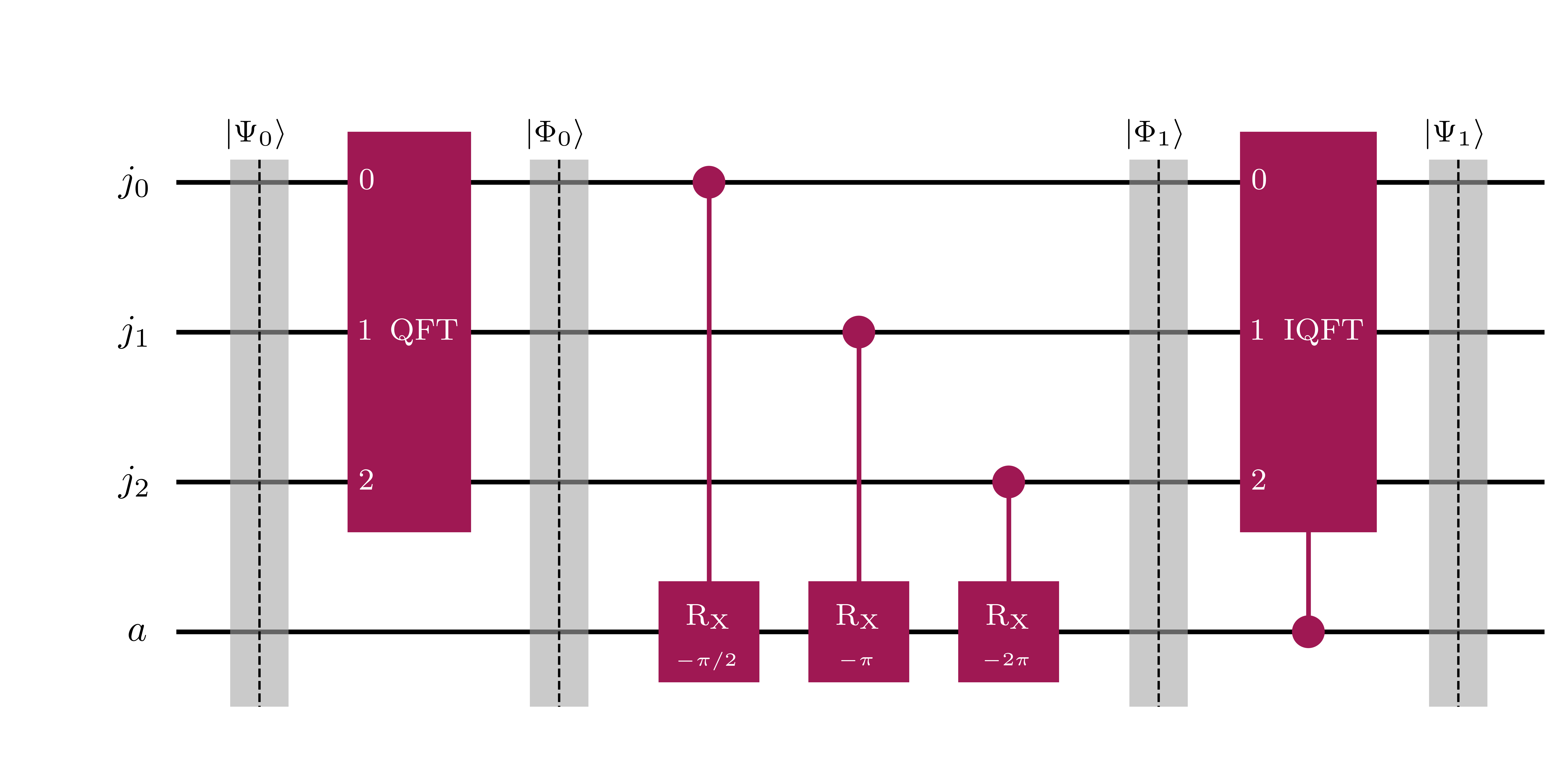}
    \caption{Three qubit example of the QFTD circuit. QFTs are applied to the $j$-register, which contains amplitude encoded function samples. Modified wavenumber scaling is accomplished by performing a series of controlled $R_x$ gates on an ancillary qubit. Following the ancilla-controlled inverse QFT operations, the QFTD states are measured.}
    \label{fig: QftdCircuitExample}
\end{figure}

Given that the modified wavenumber takes the form of a trigonometric function, element-wise multiplication between the modified wavenumber vector and the frequency space states may be performed using a sequence of controlled rotation gates.
In the following, we show the viability of this process.
First, consider the input state $\ket{\Psi_0}$ of the QFTD circuit expressed as a summation over all $N$ states of an $n$-qubit system:
\begin{equation}\label{eq: Qftd_Psi0}
    \ket{\Psi_0}=\sum_{j=0}^{N-1} \frac{f_j}{||f||_{l^2}} \ket{j} 
\end{equation}
where $||f||_{l^2}$ denotes the $l^2$ norm of the input function samples, $l^2=\sqrt{\sum_{j=0}^{N-1}|f_j|^2}$, $|f_j|$ is the magnitude of $f_j$, and the total number of samples, $N$, is given by $N=2^n$.
Here, we also define the superposition of input states as $\ket{\Psi_0}$ and use $j$ to denote individual states; thus, $f_j$ is the $j^\text{th}$ function sample which is stored in the amplitude of basis state $\ket{j}$.
Throughout the remainder of this paper, we will follow the convention that, in the input domain (position, time), superpositions of states are denoted by $\ket{\Psi}$ and individual states are indicated by $j$, while in the frequency domain, superpositions are given by $\ket{\Phi}$ and individual states are indicated by $k$.

Applying a QFT to the quantum state induces the mapping $f_j \ket{j} \rightarrow \mathcal{F}_k \ket{k} $, hence resulting in:

\begin{equation}\label{eq: Qftd_Psi1}
    \ket{\Phi_0}=
    \sum_{k=0}^{N-1} \frac{\mathcal{F}_k}{||f||_{l^2}}\ket{k}
\end{equation}

Now, we consider the desired output of the wavenumber multiplication circuit. 
From Eq.~\eqref{eq: Discrete_FiOmegaF}, we see that the $k^{\text{th}}$ state must be scaled by $i\sin(\theta_k)/\Delta x $, where $\theta_k=\frac{2\pi k}{N}$.
Noting that the $1/\Delta x$ constant may be applied during post-processing, it is temporarily neglected.
Then, the target output of the wavenumber multiplication circuit becomes:

\begin{equation}\label{eq: Qftd_phi1}
    \ket{\Phi_1} = 
    \sum_{k=0}^{N-1}i\sin(\theta_k)\frac{\mathcal{F}_k}{||f||_{l^2}}\ket{k}
\end{equation}

In order to achieve the desired form of the quantum state as in Eq.~\ref{eq: Qftd_phi1}, we choose to incorporate an ancilla qubit initialized to the $\ket{0}_a$ state, where subscript $a$ denotes the ancilla register.
We use the ancilla qubit to perform a conditional rotation of the quantum state that implements the mapping $\ket{k}\ket{0}_a \rightarrow *\ket{0}_a + i\sin(\theta_k)\ket{k}\ket{1}_a$ with the objective of isolating terms of the quantum state that did not receive a $\sin(\theta_k)$ term; here, $*$ denotes unused final states.
Given that the wavenumber introduces an imaginary number $i$, we perform the controlled rotation via a $R_x$ gate defined as:

\begin{equation}\label{eq: Rx_defn}
    R_x(\Theta) = e^{-\frac{iX}{2}\Theta} =
    \begin{bmatrix}
        \cos(\Theta/2) & -i\sin(\Theta/2) \\
        -i\sin(\Theta/2) & \cos(\Theta/2)
    \end{bmatrix}
\end{equation}
It is immediate to verify that, when this gate operates on the state $\ket{0}_a$, the resulting qubit carries an $-i\sin(\Theta/2)$ coefficient on the $\ket{1}_a$ state.
Indeed, assuming a rotation of magnitude $-2\Theta$ and leveraging the parity properties of the trigonometric functions, the gate transforms the state $\ket{0}_a$ as:

\begin{equation}\label{eq: Rx_on_1}
    R_x(-2\Theta)\ket{0}_a = 
    \cos(\Theta)\ket{0}_a + i\sin(\Theta)\ket{1}_a
\end{equation}
Therefore, we seek to perform the following transformation:

\begin{equation}\label{eq: WavenumberRotationGoal}
    \sum_{k=0}^{N-1} \frac{\mathcal{F}_k}{||f||_{l^2}} \ket{k}\ket{0}_a 
    \rightarrow 
    \sum_{k=0}^{N-1} \bigg[
    \cos(\theta_k) \frac{\mathcal{F}_k}{||f||_{l^2}} \ket{k} \ket{0}_a +  i\sin(\theta_k) \frac{\mathcal{F}_k}{||f||_{l^2}} \ket{k} \ket{1}_a \bigg]
\end{equation}
or equivalently: 

\begin{equation}
    \sum_{k=0}^{N-1} \frac{\mathcal{F}_k}{||f||_{l^2}} \ket{k} \ket{0}_a
    \rightarrow 
    \sum_{k=0}^{N-1} \frac{\mathcal{F}_k}{||f||_{l^2}} \ket{k}e^{-\frac{iX}{2}(-2\theta_k)}\ket{0}_a
    \
\end{equation}

At this point, the remaining description of the modified wavenumber scaling follows closely from the proof of the QFT algorithm.
As such, we now express the argument to the $R_x$ gate, $-2\theta_k$, in its binary form:
\begin{equation}\label{eq: theta_binary}
    \begin{split}
    -2\theta_k & = -2\bigg[\frac{2\pi}{2^n}(2^{n-1}k_{n-1} + 2^{n-2}k_{n-2} + ... + 2k_1 +k_0)\bigg]\\
    & = -\pi\big(2^1k_{n-1} + 2^{0}k_{n-2} + ... + 2^{3-n}k_1 +2^{2-n}k_0\big)
    \end{split}
\end{equation}
and our desired state becomes: 
\begin{equation}\label{eq: LongExp}
    \ket{\Phi_1} = 
    \sum_{k=0}^{N-1} \frac{\mathcal{F}_k}{||f||_{l^2}} \ket{k}
    \exp\bigg[-\frac{iX}{2}\bigg(
    -\pi\big(2^1k_{n-1} + 2^{0}k_{n-2} + ... + 2^{3-n}k_1 +2^{2-n}k_0\big)
    \bigg)\bigg]
    \ket{0}_a
    \
\end{equation}

To observe how gates must be implemented at the individual qubit level, we wish to express $\ket{\Phi_1}$ in its product state.
Because $\mathcal{F}_k$ varies with each state, it must be written in a different form that may be factored into the product state.
To that end, we define $\mathcal{F}_k$ as the product:
\begin{equation}\label{eq: ProductF}
    \mathcal{F}_k = 
    b_{k_{n-1}}^{(n-1)} b_{k_{n-2}}^{(n-2)} \cdots b_{k_1}^{(1)} b_{k_0}^{(0)}
    \,,
\end{equation}
where the superscript of each $b$ coefficient denotes a particular qubit, and the subscript denotes that qubit's $\ket{0}$ or $\ket{1}$ state. For example, if $k=1$, then $\mathcal{F}_k$ is the coefficient of the state $\ket{00\cdots01}$, and $\mathcal{F}_k = b_0^{(n-1)} b_0^{(n-2)} \cdots b_0^{(1)} b_1^{(0)}$.
Using this notation to factor $\mathcal{F}_k$ while writing $\ket{k}$ as a tensor product of the $n$ qubits, $\ket{k_{n-1}}\ket{k_{n-2}}\cdots\ket{k_1}\ket{k_0}$, and rearranging the exponential terms of Eq.~\eqref{eq: LongExp}, $\ket{\Phi_1}$ may be expressed in its product state:
\begin{equation}\label{eq: Qftd_ProductState}
\begin{split}
    \ket{\Phi_1} =\frac{1}{||f||_{l^2}} &
    \bigg[ \sum_{k_{n-1}=0}^{1} 
    b_{k_{n-1}}^{(n-1)} 
    e^{-\frac{i X}{2} (-2^1 \pi k_{n-1})} 
    \ket{k_{n-1}}\bigg]
    \otimes 
    \bigg[ \sum_{k_{n-2}=0}^1 
    b_{k_{n-2}}^{(n-2)} 
    e^{-\frac{i X}{2} (-2^{0} \pi k_{n-2})}
    \ket{k_{n-2}}\bigg]
    \otimes \cdots \\
    & \cdots \otimes
    \bigg[ \sum_{k_1=0}^1
    b_{k_1}^{(1)} 
    e^{-\frac{i X}{2} (-2^{3-n} \pi k_1)}
    \ket{k_1} \bigg]
    \otimes
    \bigg[ \sum_{k_0=0}^1
    b_{k_0}^{(0)} e^{-\frac{i X}{2}(- 2^{2-n} \pi k_0)}\ket{k_0}\bigg]
    \otimes \ket{0}_a
\end{split}
\end{equation}
Then, the final product state is obtained by evaluating each summation:

\begin{equation}\label{eq: Qftd_FinalProductState}
\begin{split}
    \ket{\Phi_1} = \frac{1}{||f||_{l^2}} &
    \bigg[b_0^{(n-1)} \ket{0} + b^{(n-1)}_1 e^{-\frac{i X}{2} (-2^1 \pi)} \ket{1}\bigg] \otimes
    \bigg[b_0^{(n-2)} \ket{0} + b_1^{(n-2)} e^{-\frac{i X}{2}(- 2^{0} \pi)}\ket{1}\bigg]\otimes
    \cdots \\ & \cdots \otimes
    \bigg[b_0^{(1)} \ket{0} + b_1^{(1)}e^{-\frac{i X}{2}(- 2^{3-n} \pi)}\ket{1}\bigg]\otimes
    \bigg[b_0^{(0)}\ket{0} + b_1^{(0)} e^{-\frac{i X}{2}(- 2^{2-n} \pi)}\bigg]\otimes
    \ket{0}_a
    \,.
\end{split}
\end{equation}

\begin{figure}[!h]
    \centering
    \includegraphics[width=0.25\linewidth]{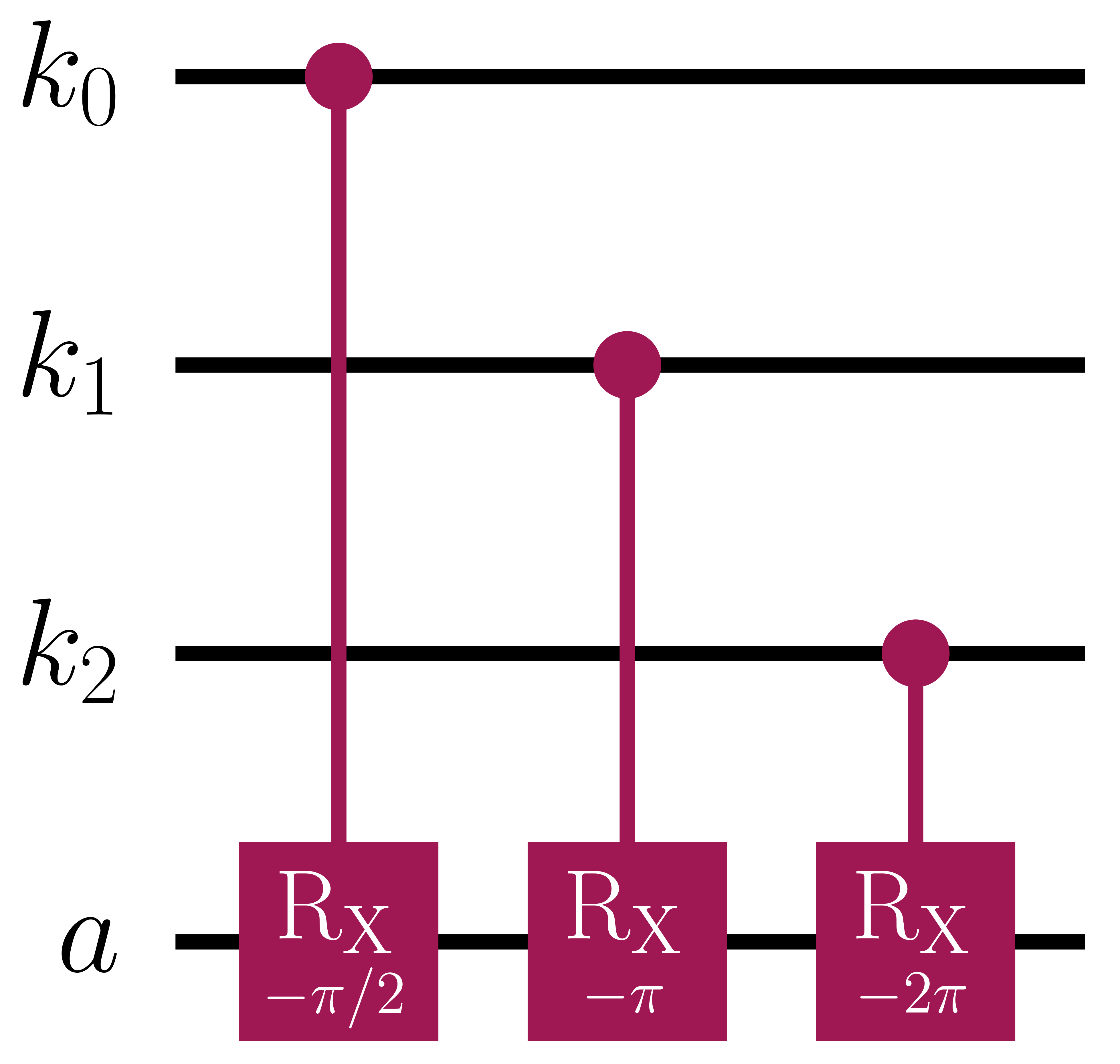}
    \caption{Three qubit example of the wavenumber rotation portion of the QFTD circuit. Controlled rotation gates transform the ancilla to the state $\ket{0}_a$ when the $k^{\text{th}}$ superposition state is scaled by $i\sin(2\pi k /N)$.}
    \label{fig: WavenumberRotationCirc}
\end{figure}

As $\ket{\Phi_0}$ is an entangled state, it is important to recognize that the $b$ coefficients appearing in Eq.~\eqref{eq: Qftd_FinalProductState} cannot be algebraically obtained \cite{wong_introduction_2022, nielsen_quantum_2010}; however, determining their exact values is not needed in the present approach.
Focusing on the exponential terms in Eq.~\eqref{eq: Qftd_FinalProductState}, it is observed that exponential terms only appear when the $p^{th}$ qubit is in state $\ket{1}$, that is $\ket{k_p}=\ket{1}$. It follows that the $k$-qubits can serve as controls for a sequence of ancillary qubit rotations in order to generate the correct final state.
Furthermore, we see that the $p^{th}$ qubit in the $k$-register must control an ancilla rotation of $\phi_p = -2^{p-n+2}\pi$ radians.
Accordingly, the wavenumber rotation circuit presented in Fig.~\ref{fig: WavenumberRotationCirc} shows the $p^{\text{th}}$ qubit controlling an $R_x$ gate with the argument $-2^{p-n+2}\pi$ radians.

To illustrate more clearly the effects of the controlled rotations, we restate Eq.~\eqref{eq: Qftd_FinalProductState} using the associative property of the Kronecker product, that is $\ket{a} \otimes \ket{b} \otimes \ket{c} = \big(\ket{a} \otimes \ket{b}\big) \otimes\ket{c}$. In the proceeding expression, note that the ancilla qubit is only subjected to a rotation of an angle $\phi_p$ when the $p^{th}$ qubit is in state $\ket{1}$: 

\begin{equation}\label{eq: RecombinedTensor1}
    \ket{\Phi_1} = 
    \frac{1}{||f||_{l^2}}
    \begin{pmatrix}
        b_0^{(n-1)} b_0^{(n-2)} \cdots b_0^{(1)} b_0^{(0)}
        \\
        b_0^{(n-1)} b_0^{(n-2)} \cdots b_1^{(1)} b_0^{(0)} ~
        e^{-\frac{i X}{2} \phi_0} 
        \\
        b_0^{(n-1)} b_0^{(n-2)} \cdots b_1^{(1)} b_0^{(0)} ~
        e^{-\frac{i X}{2} \phi_1} 
        \\
        b_0^{(n-1)} b_0^{(n-2)} \cdots b_1^{(1)} b_1^{(0)} ~
        e^{-\frac{i X}{2} (\phi_1 + \phi_0)} 
        \\
        \vdots \\
        b_1^{(n-1)} b_1^{(n-2)} \cdots b_1^{(1)} b_1^{(0)} ~
        e^{-\frac{i X}{2} (\phi_{n-1} + \phi_{n-2} + \cdots +\phi_1 + \phi_0)} \\
    \end{pmatrix}
    \otimes \ket{0}_a 
\end{equation}
Referring to $\theta_k$'s binary form in Eq.~\ref{eq: theta_binary}, we may express $-2\theta_k$ as a function of the controlled rotation angles:
\begin{equation}\label{eq: ThetaOfPhi}
    -2\theta_k = 
    \phi_{n - 1}  k_{n - 1} + 
    \phi_{n - 2}k_{n - 2} +\cdots + 
    \phi_1k_1 + 
    \phi_0k_0
\end{equation}
Combining Eq.~\eqref{eq: ProductF}, Eq.~\eqref{eq: RecombinedTensor1} and Eq.~\eqref{eq: ThetaOfPhi}, we perform the following simplification:
\begin{equation}\label{eq: RecombinedTensor2}
    \ket{\Phi_1} =
    \frac{1}{||f||_{l^2}}
    \begin{pmatrix}
        \mathcal{F}_0 e^{-\frac{i X}{2}(-2 \theta_0)} \\
        \mathcal{F}_1 e^{-\frac{i X}{2}(-2 \theta_1)} \\
        \mathcal{F}_2 e^{-\frac{i X}{2}(-2 \theta_2)} \\
        \mathcal{F}_3 e^{-\frac{i X}{2}(-2 \theta_3)} \\
        \vdots \\
        \mathcal{F}_{N - 1} e^{-\frac{i X}{2}(-2 \theta_{N - 1})} \\
    \end{pmatrix}
    \otimes
    \begin{pmatrix}
        1\\
        0
    \end{pmatrix}_a
\end{equation}
Finally, the last Kronecker product is evaluated and the $R_x$ operators included in the $k$ vector are allowed to act on the ancilla qubit:

\begin{equation}\label{eq: FishyOperatorNotation}
    \ket{\Phi_1} = 
    \frac{1}{||f||_{l^2}}
    \begin{pmatrix}
        \mathcal{F}_0 e^{-\frac{i X}{2} (-2\theta_0)} 
        \begin{psmallmatrix}
            1\\
            0
        \end{psmallmatrix}_a\\
        \mathcal{F}_1 e^{-\frac{i X}{2} (-2\theta_1)}
        \begin{psmallmatrix}
            1\\
            0
        \end{psmallmatrix}_a\\
        \mathcal{F}_2 e^{-\frac{i X}{2} (-2\theta_2)}
        \begin{psmallmatrix}
            1\\
            0
        \end{psmallmatrix}_a\\
        \mathcal{F}_3 e^{-\frac{i X}{2} (-2\theta_3)}
        \begin{psmallmatrix}
            1\\
            0
        \end{psmallmatrix}_a\\
        \vdots \\
        \mathcal{F}_{N \text{-} 1} e^{-\frac{i X}{2} (-2\theta_{N \text{-} 1})}
        \begin{psmallmatrix}
            1\\
            0
        \end{psmallmatrix}_a\\
    \end{pmatrix}
    =
    \begin{pmatrix}
         \cos(\theta_0) \mathcal{F}_0 \\
         \cos(\theta_1) \mathcal{F}_1 \\
         \cos(\theta_2) \mathcal{F}_2 \\
         \cos(\theta_3) \mathcal{F}_3 \\
        \vdots\\
        \cos(\theta_{N \text{-} 1}) \mathcal{F}_{N \text{-} 1}
    \end{pmatrix}
    \otimes \ket{0}_a +
    \begin{pmatrix}
        i\sin(\theta_0) \mathcal{F}_0 \\
        i\sin(\theta_1) \mathcal{F}_1 \\
        i\sin(\theta_2) \mathcal{F}_2 \\
        i\sin(\theta_3) \mathcal{F}_3 \\
        \vdots\\
        i\sin(\theta_{N \text{-} 1}) \mathcal{F}_{N \text{-} 1}
    \end{pmatrix}
    \otimes \ket{1}_a 
\end{equation}
Noting that the $\sin(\theta_k)$ terms only appear when the ancilla qubit is in the state $\ket{1}_a$, Eq.~\eqref{eq: FishyOperatorNotation} reveals that we have successfully obtained a state in the form of Eq.~\eqref{eq: WavenumberRotationGoal}.  Note, the ancilla qubit was included in the LSB position while presenting the underlying theory of the QFTD circuit; however, the ancilla is encoded in the most significant bit (MSB) position in all circuit examples so that states receiving the proper modified wavenumber are easily identified when viewing the full set of circuit observations.

Once the $\ket{\Phi_1}$ is obtained, the algorithm is completed by applying an inverse QFT to the $k$-register. Given that only the $\ket{1}_a$ states have received the correct phase, the inverse QFT must be controlled by the $\ket{1}_a$ state. Again denoting unused states with $*$, the ancilla-controlled inverse QFT casts the system into the following state:

\begin{equation}
    \ket{\Psi_1} = *\ket{0}_a + QFT^{-1}\Bigg[\sum_{k=0}^{N-1} i\sin(\theta_k) \mathcal{F}_k \ket{k}\ket{1}_a\Bigg]
\end{equation}

Following the controlled inverse-QFT, $\ket{\Psi_1}$ is proportional to the numerical derivative of the input function when the ancilla qubit is in state $\ket{1}_a$.
After measuring all qubits, observations with $\ket{0}_a$ are discarded and the correct amplitudes are recovered through post-processing. 
The input function's norm and the constant multiplier $1 / \Delta x$ must be factored classically to recover the amplitude. 
Finally, given that each state's probability amplitude is proportional to the squared state coefficient, the measured values are proportional to the square of the numerical derivative. Therefore, the correct amplitude is recovered by:

\begin{equation}\label{eq: Qftd_AmplitudeRecovery}
    \bigg[\frac{\Delta f}{\Delta x}\bigg]_j^2 = \bigg(\frac{||f||_{l^2}}{\Delta x}\bigg)^2 ~\psi_j^2
    \,,
\end{equation}
where $\psi_j^2$ represents the squared amplitude of the $j^{th}$ state. Using Eq.~\eqref{eq: Qftd_AmplitudeRecovery}, the minimum resolution of the QFTD algorithm may be obtained. For a circuit simulated using $M$ shots, the measurement least count is $\psi^2 = 1/M$. Substituting into Eq.~\eqref{eq: Qftd_AmplitudeRecovery}, the QFTD resolution is obtained:
\begin{equation}\label{eq: Qftd_Resolution}
    r = \frac{||f||_{l^2}^2}{M\Delta x^2}
\end{equation}
where $r$ denotes the QFTD algorithm's resolution; \textit{not} the error associated with the QFTD algorithm, $\epsilon$.
The resolution, or \enquote{least count}, of the algorithm exactly defines its smallest non-zero output.
Furthermore, all non-zero outputs of the QFTD are rounded to integer multiples of $r$. As such, in regions where the square of the input function's derivative is less than $r/2$, the associated states are not expected to be observed during measurement.
At such points, a value of zero is assumed for the QFTD output during post-processing.

\subsection{The Quantum Fourier Transform Integration (QFTI)}

This section presents the techniques necessary for a quantum implementation of the DFT-based integration algorithm described above. Particular attention is given to 1) the application of the $\cos(\omega_k\Delta x)$ terms of the QFTI modified wavenumber, and 2) the derivation of a unitary block encoding of the partial summation matrix, $\Sigma$, included in Eq.~\eqref{eq: LowerTri_mat}.

A direct comparison of Eq.~\eqref{eq: Discrete_FiOmegaF} and Eq.~\eqref{eq: DiffArea} reveals that the differential and integral modified wavenumber formulations differ only by a constant multiplier and by the presence of a phase term.
This observation allows using an identical wavenumber rotation circuit to scale each term with the appropriate modified wavenumber. 
Referring to the definition of the $R_x$ gate in Eq.~\eqref{eq: Rx_defn}, scaling by the term $\cos(\frac{2\pi k}{N})$ is accomplished when the series of ancilla rotations result in an ancilla state of $\ket{0}_a$.
Then, upon measuring the final output, observations with the ancilla qubit in the state $\ket{0}_a$ have received the proper phase, while observations in the $\ket{1}_a$ should be discarded.

Unlike the QFTD algorithm, the QFTI algorithm's application of the QFT, the wavenumber rotation, and the inverse-QFT only encodes a vector of the differential areas of the function $f$.
Shown in Eq.~\eqref{eq: LowerTri_mat}, the integral approximation is accomplished by applying the unit lower left matrix, $\Sigma$, to the quantum statevector containing differential area estimates.
To incorporate this matrix product operation, we seek to encode the matrix $\Sigma$ within a larger unitary matrix which may be initialized in a quantum circuit as an arbitrary unitary gate.

To construct the unitary matrix that applies $\Sigma$ to the quantum statevector, we construct a Hermitian matrix, $H$, as follows:
\begin{equation}
    H = 
    \begin{bmatrix}
        0 & \Sigma^\dagger \\
        \Sigma & 0
    \end{bmatrix}
\end{equation}
where the symbol $\dagger$ denotes the complex conjugate transpose operation. Following the procedure for block encoding of a matrix product operator \cite{nibbi_block_2024}, it is possible to embed the hermitian matrix $H$ into a unitary matrix $U_H$ having the form:
\begin{equation}\label{eq: Uh}
    U_H = 
    \begin{bmatrix}
        H/\eta & C \\
        B & D
    \end{bmatrix}
\end{equation}
where $B$, $C$, and $D$ are particular matrices to be identified, and $\eta$ is the spectral norm of $H$ which is defined to be the largest eigenvalue of the matrix $\sqrt{HH^\dagger}$.
Given that $U_H$ is defined to be unitary, $B$ must be a matrix such that $BB^\dagger = I - HH^\dagger$\cite{nibbi_block_2024}, where $I$ is the identity matrix.
Expressing $H$ as the product of a diagonal matrix $D$ and its eigenvector matrices $P$ and $P^\dagger$, we obtain:
\begin{equation}
    H = PDP^\dagger
\end{equation}

The expression for $BB^\dagger$ may be written in terms of known quantities; therefore, $B$ may be found by evaluating:
\begin{equation}
    B=\sqrt{I-D^2/\eta^2}P^\dagger
\end{equation}
Finally, defining the rectangular matrix $W=\begin{bsmallmatrix} H/\eta \\B \end{bsmallmatrix}$ and performing QR-decomposition allows $W$ to be written as the product of the unitary matrix $Q$, and the rectangular matrix $R$:
\begin{equation}
    \begin{split}
        W &= QR \\
        & = U_H R
    \end{split}
\end{equation}
where $Q$ is a matrix of the form $U_H$ given in Eq.~\eqref{eq: Uh}.
Note that the QR decomposition of $W$ is not unique.
In fact, given that both $W=QR$ and $W=(-Q)(-R)$, it must be enforced that $U_H$ is a block encoding of $\Sigma$ rather than $-\Sigma$, scaling $U_H$ by $-1$ as needed.

\begin{figure}[h!]
    \centering
    \includegraphics[width=0.85\linewidth]{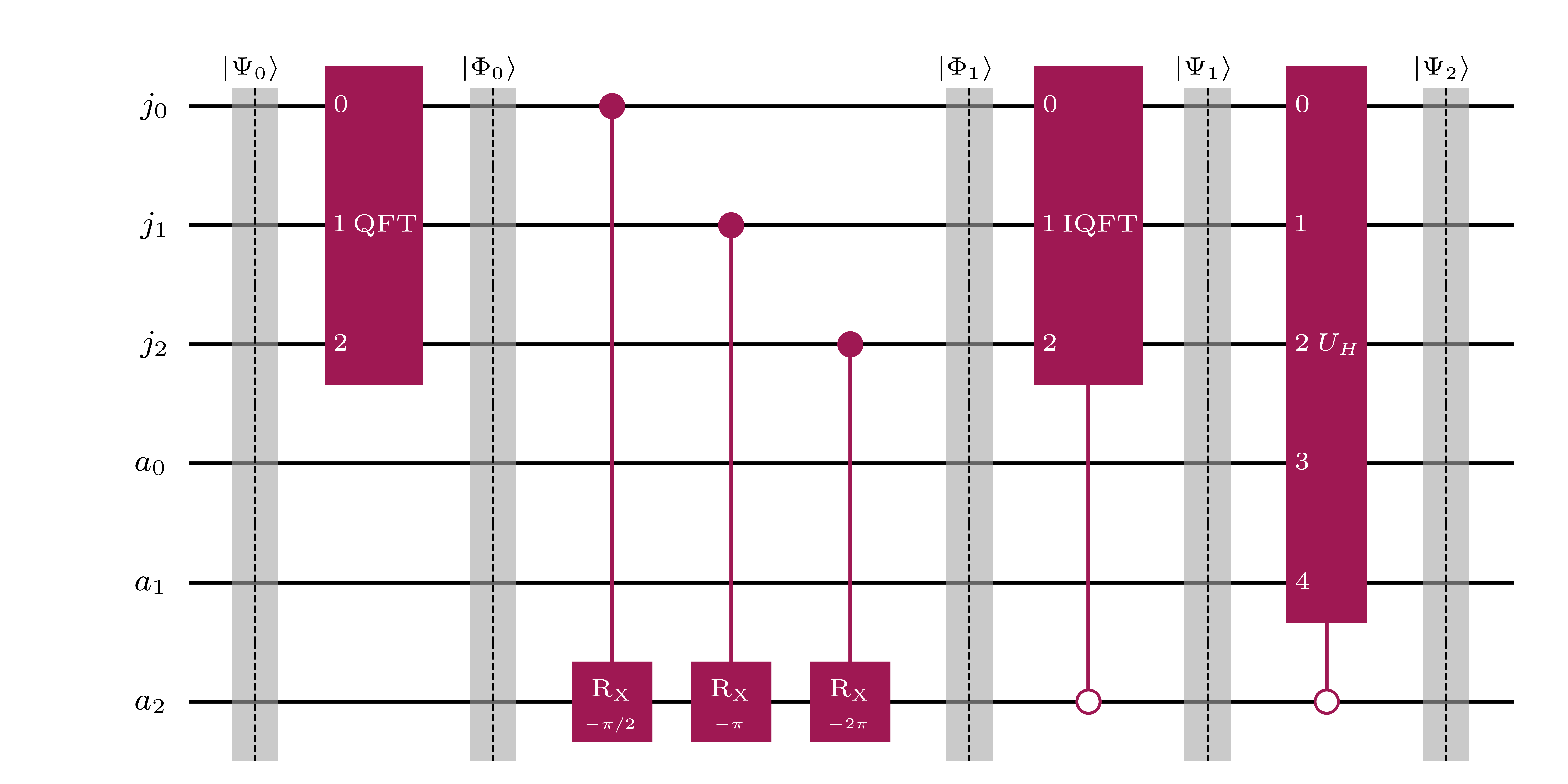}
    \caption{Three qubit example of the QFTI circuit. Following a similar procedure used in the QFTD algorithm, the differential area of the input function is encoded for each point. Partial summations of each differential area are performed using the partial summation matrix product operator (PsMPO) gate to approximate the integral at each point.}
    \label{fig: QftiCircuit}
\end{figure}

In the QFTI circuit shown in Fig.~\ref{fig: QftiCircuit}, and in the QFTI examples to follow, we perform the matrix product operation $U_H\ket{\Psi_1}$ by initializing $U_H$ as an arbitrary unitary gate.
This is accomplished using the built-in \enquote{Unitary Gate} encoding function available in Qiskit. 
This function employs decomposition algorithms\cite{iten_quantum_2016} to construct circuits corresponding to arbitrary isometries.
In the case of $n \times n$ unitary matrices, the resulting circuits have gate-level complexities ranging from $\frac{1}{4}4^n$ to $\frac{23}{48}4^n$. 
Different encoding strategies could certainly be envisioned depending on specific requirements and target performance.
However, this approach shows that even a classical encoding strategy works well with our proposed method.
The topic of block matrix encoding is in itself a highly active research area that has already produced several techniques, which improve upon the gate-level encoding complexities\cite{camps_explicit_2024, van_dam_simulating_2011, low_hamiltonian_2019, yosef_encoding_2025}.
These are computational improvements that could be explored and integrated in future work, but they will not alter the core approach presented here for differentiation and integration.

While obtaining the unitary summation matrix, $U_H$, the overall size of the problem was increased multiple times.
Starting from the lower-triangular matrix $\Sigma$, the operator's size was doubled to construct the hermitian matrix $H$.
While constructing $U_H$, the size of the partial sum operator was doubled again.
In total, the block encoding of $U_H$ calls for the inclusion of two additional ancilla qubits in the QFTI circuit.
In Fig.~\ref{fig: QftiCircuit}, the two additional qubits are encoded in the LSB positions of the 3-qubit $a$ register and initialized to $\ket{0}_{a_1}$ and $\ket{0}_{a_0}$.
To ensure the summation operation only includes states which received the appropriate phase during the wavenumber rotations, we control the summation operation using the $\ket{0}_{a_2}$ state of the $a_2$ qubit which was used for performing wavenumber rotations.
Then, the input to the partial summation matrix product gate is $\ket{0}_{a_1}\ket{0}_{a_0}\ket{\Psi_1}$.
Performing the controlled $U_H$ operation, we find the state $\ket{\Psi_2}$ to be:
\begin{equation}
    \ket{\Psi_2}= 
    \sum_{j=0}^{N-1}\ket{0}_{a_2}\big[U_H(\ket{00}\otimes A_j\ket{j})\big]+*\ket{1}_{a_2}
    = 
    \ket{0}_{a_2}
    \begin{bmatrix}
        0 & \Sigma^\dagger/\eta & * & * \\
        \Sigma/\eta & 0 & * & * \\
        * & * & * & * \\
        * & * & * & *
    \end{bmatrix}
    \begin{pmatrix}
        \vec{A} \\
        \vec{0} \\
        \vec{0} \\
        \vec{0}
    \end{pmatrix}
    + *\ket{1}_{a_2}
\end{equation}
where $\vec{A}$ denotes the vector of differential areas encoded in the amplitudes of $\ket{\Phi_1}$, the $B$, $C$, and $D$ are replaced using placeholder $*$ notation, and zero vectors padding $\vec{A}$ to the size of $U_H$ are denoted by $\vec{0}$.
Evaluating the matrix product and Kronecker product, the final state is obtained:
\begin{equation}
    \ket{\Psi_2} =
    \begin{pmatrix}
        \vec{0} \\
        \Sigma \vec{A}/\eta \\
        * \\
        * \\
        * \\
        \vec{0} \\
        \vec{0} \\
        \vec{0}
    \end{pmatrix}
\end{equation}

From the final expression of $\ket{\Psi_2}$, it is observed that, for the little endian bit-ordering of this work, the partial summation results are stored in states where the three most significant bits are $\ket{001}$. At this time, the coefficients of the $\ket{001\cdots}_a\ket{j}$ states are proportional to $I_j$.

As was the case for the QFTD, upon measuring the statevector, the results must be scaled to recover the correct amplitudes. Referencing Eq.~\eqref{eq: DiffArea}, we see the coefficients must be scaled by a factor of $1/\Delta x$ to account for its absence during the wavenumber rotation procedure.
Additionally, the norm of the input function $||f||_{l^2}$, and the spectral norm of the summation matrix $\eta$ must also be factored into the result. Finally, we recall that the measured quantities are not the state coefficients but their squared values. It follows that the circuit output is given by:
\begin{equation}\label{eq: QFTI_AmplitudeRecovery}
    \big[I_j\big]^2 = \big(||f||_{l^2} \eta  \Delta x\big)^2\psi_j^2
\end{equation}
Again, the QFTI algorithm's resolution for a simulation using $M$ shots is obtained by evaluating Eq.~\eqref{eq: QFTI_AmplitudeRecovery} at $\psi^2 = 1/M$:
\begin{equation}\label{eq: QFTI_Resolution}
    r = \frac{\big(||f||_{l^2} \eta \Delta x\big)^2}{M}
\end{equation}
As before, non-zero QFTI outputs are rounded to the nearest integer multiple of $r$; and, in regions where the analytical output is lesser than the $r/2$, the associated states are not expected to be observed when measuring the circuit. Therefore, when a state is not observed its QFTI output is assigned a value of zero during post-processing.

\subsection{Extensions to Higher Dimensions}
In this section, we discuss how the QFTD algorithm can be extended to applications in higher-dimensions.
Noting that the partial summation approximation of an integral employed by the QFTI does not extend to higher-dimensional integrals, we do not consider higher-dimensional QFTI extensions in this work.
In the case of the QFTD algorithm, the higher-dimensional implementation bears similarities with its one-dimensional counterpart with exceptions including the input encoding, parallel QFTD applications, and the scope of results.

\begin{figure}[!h]
    \centering
    \includegraphics[width=.85\textwidth]{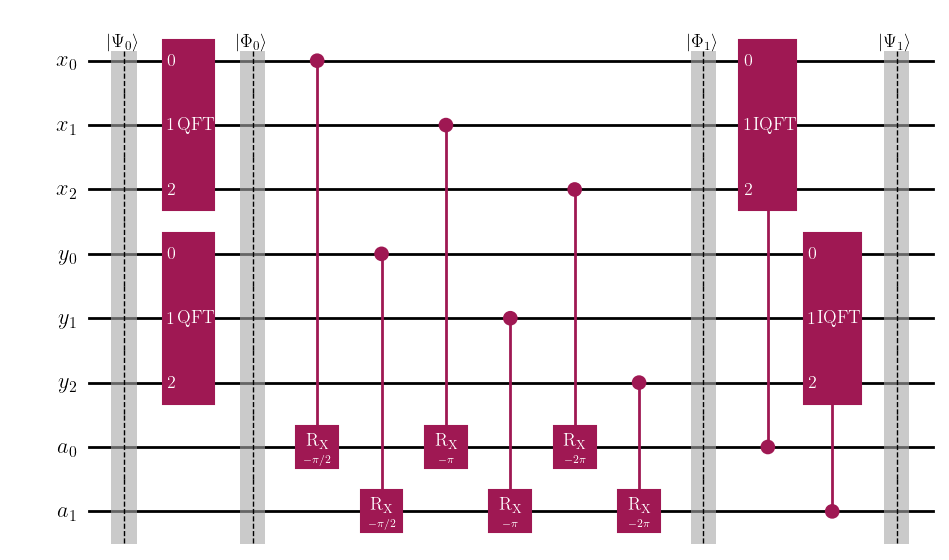}
    \caption{Two-dimensional QFTD circuit for input discretized by $n=3$ qubits in each dimension.}
    \label{fig: ThreeQubit_NdQFTD_Circ}
\end{figure}

First, the input function must be encoded carefully to ensure partial derivatives may be evaluated separately.
To illustrate our encoding scheme, we consider a generalized $d$-dimensional domain, where the $p^\text{th}$ dimension is discretized by $n_p$ qubits into $N_p=2^{n_p}$ points.
To prepare this input appropriately for the multidimensional extension of the QFTD, function samples must be encoded in $d$ registers, where the register corresponding to the $p^\text{th}$ dimension, $\ket{x^{(p)}}$, contains $n_p$ qubits.
For illustration, general $d$-dimensional input function samples are encoded according to Eq.~\eqref{eq: ndEncoding}.
For the specific case of a two-dimensional QFTD operation, such as that illustrated by Fig.~\ref{fig: ThreeQubit_NdQFTD_Circ}, the encoding scheme is presented again in \S~\ref{sec: ndQFTD_Examples} alongside an example of a two-dimensional QFTD application.
\begin{equation}\label{eq: ndEncoding}
    \ket{\Psi_0} = 
    \ket{x^{(d-1)}}^{\otimes{n_{d-1}}} \otimes 
    \cdots \otimes
    \ket{x^{(1)}}^{\otimes{n_1}} \otimes
    \ket{x^{(0)}}^{\otimes{n_0}}
    = 
    \begin{pmatrix}
        f\big(x^{(d-1)}_0,~\cdots,~x^{(1)}_0,~x^{(0)}_0\big)\\
        f\big(x^{(d-1)}_0,~\cdots,~x^{(1)}_0,~x^{(0)}_1\big)\\
        \vdots \\
        f\big(x^{(d-1)}_0,~\cdots,~x^{(1)}_0,~x^{(0)}_{N_0-1}\big)\\
        f\big(x^{(d-1)}_0,~\cdots,~x^{(1)}_1,~x^{(0)}_0\big)\\
        \vdots \\
        f\big(x^{(d-1)}_{N_d-1},~\cdots,~x^{(1)}_{N_1-1},~x^{(0)}_{N_0-2}\big)\\
        f\big(x^{(d-1)}_{N_d-1},~\cdots,~x^{(1)}_{N_1-1},~x^{(0)}_{N_0-1}\big)
    \end{pmatrix}
\end{equation}

Second, parallel implementations of the one-dimensional QFTD procedure must be applied to each dimension's register. Then, final result is obtained by the tensor product of each dimension's register.
Previous studies have already proven that a multidimensional QFT is achieved by performing one-dimensional QFT operations on parallel registers corresponding to each dimension \cite{pfeffer_multidimensional_2023}.
Noting that the wavenumber application procedure follows the same structure as the QFT, the multidimensional wavenumber application follows the multidimensional QFT procedure.
To illustrate this procedure, Figure~\ref{fig: ThreeQubit_NdQFTD_Circ} illustrates a two-dimensional QFTD circuit, where each dimension is discretized by $3$ qubits to form a $2^3 \times2^3$ grid.

Third, this extension of the QFTD algorithm yields \textit{all} partial derivatives of the input function; and, the type of partial derivative contained in the primary register is indicated by ancilla qubit states.
In our interpretations of the ancilla states, we assume the $k^\text{th}$ ancilla qubit, $\ket{a^{(k)}}$, is used to perform the QFTD on the $\ket{x^{(k)}}$ register.
If $\ket{a^{(k)}}$ is found to be in the $\ket{1}$ state, the input function has been partially differentiated with respect to $x^{(k)}$.
If $\ket{a^{(k)}}$ is found to be in the $\ket{0}$ state, then no partial derivatives are performed with respect to $x^{(k)}$.
Furthermore, this reasoning also extends to mixed partial derivatives: if all ancilla qubits are in the state $\ket{1}$, the input has been partially differentiated with respect to each dimension.
Likewise, if no ancilla qubit is in the $\ket{1}$ state, the multidimensional QFTD algorithm failed to perform any partial derivatives, and the associated results are discarded.
Table \ref{table: MixedPartials} summarizes all possible two-dimensional QFTD results using subscript notation to indicate partial derivatives.
\begin{table}[h!]
\center
\begin{tabular}{c|c}
    \textbf{Ancilla State} & \textbf{Output}\\ 
    \hline
    $\ket{00}$ & Operation Failed\\
    \hline
    $\ket{01}$ & $f_{x^{(0)}}$\\
    \hline
    $\ket{10}$ & $f_{x^{(1)}}$\\
    \hline
    $\ket{11}$ & $f_{x^{(1)} x^{(0)}}$\\
\end{tabular}
\caption{Summary of two-dimensional QFTD ancilla states and corresponding algorithm output.}
\label{table: MixedPartials}

\end{table}

\subsection{Recovery of Signed Quantities}
Upon completion of the QFTD or QFTI procedures, the signed derivative or integral results are successfully encoded within the quantum state amplitudes; this information is available to proceeding computations. 
However, direct measurement of the quantum state vectors yields the square of each state amplitude, therefore not allowing to access the sign information. 
In this section, we present a procedure that may be employed to recover the sign information upon measurement.

\begin{figure}[!h]
    \centering
    \includegraphics[width=0.5\textwidth]{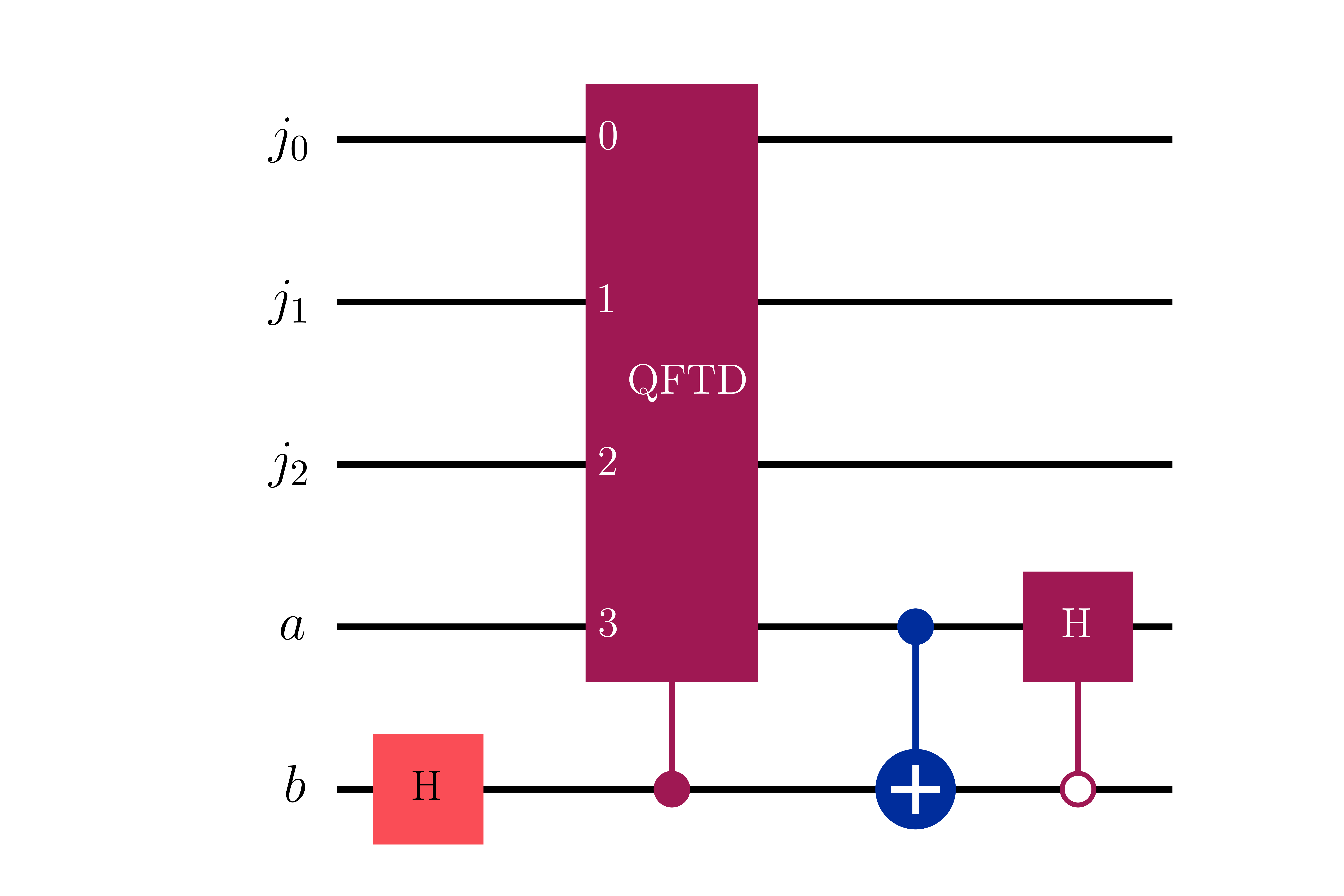}
    \caption{Example sign recovery circuit used to recover the signed output of a three-qubit QFTD operation.}
    \label{fig: Qftd_SignRecoveryCirc}
\end{figure}

\begin{figure}[!h]
    \centering
    \includegraphics[width=0.5\textwidth]{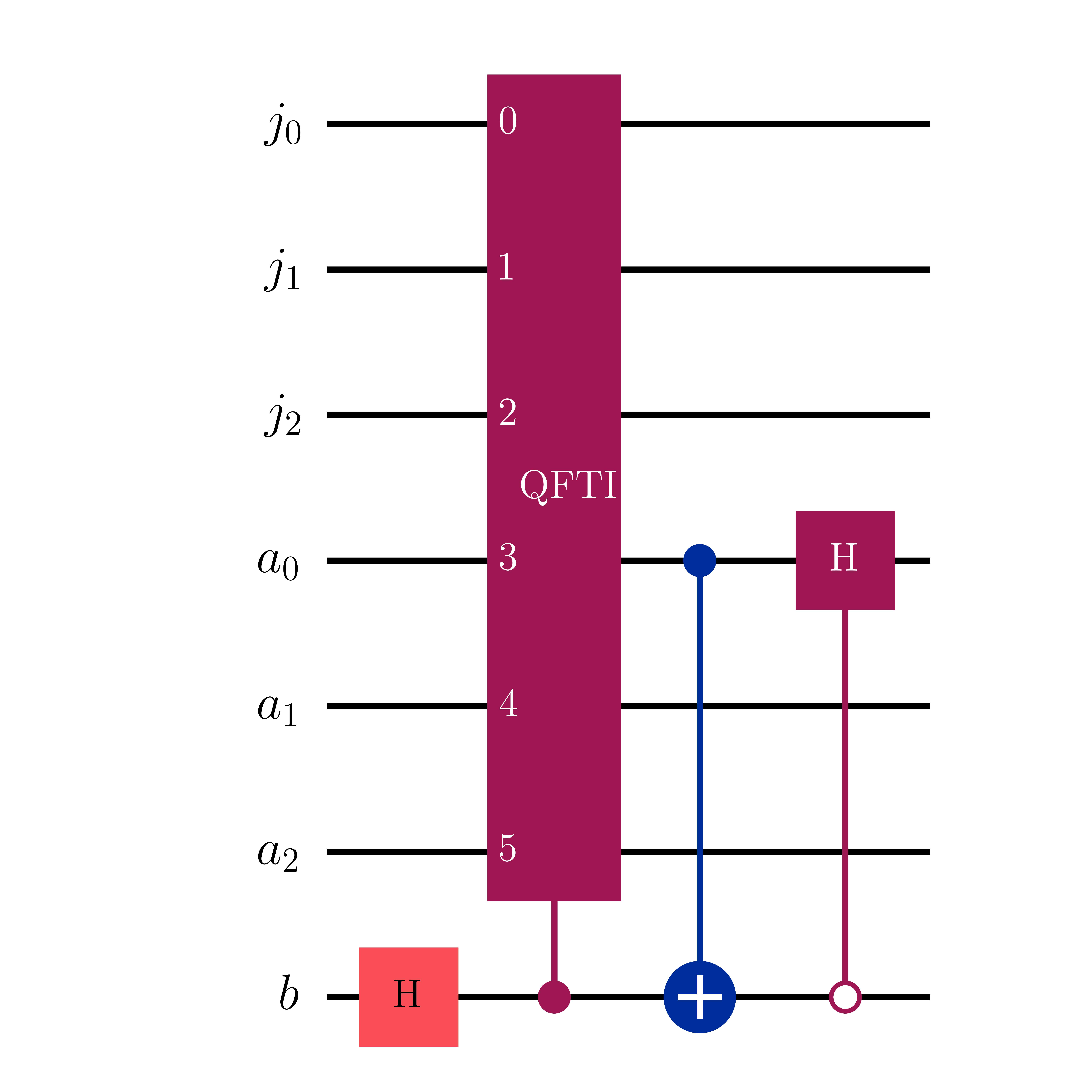}
    \caption{Example sign recovery circuit used to recover the signed output of a three-qubit QFTI operation.}
    \label{fig: QFTI_SignRecoveryCirc}
\end{figure}

We begin by noting that the sign recovery circuit and corresponding post-processing procedure \textit{only} yields the sign information of the derivative or integral results.
Thus, the previously presented QFTD must be simulated in parallel with the sign recovery circuit shown in Fig.~\ref{fig: Qftd_SignRecoveryCirc} to determine the full (signed) result. Similarly, the QFTI circuit must be simulated in parallel with the QFTI sign recovery circuit shown in Fig.~\ref{fig: QFTI_SignRecoveryCirc}.
Given that both algorithms follow nearly identical sign recovery procedures, we begin with an in-depth discussion of the QFTD sign recovery procedure and conclude with a brief description of the modifications needed for its application to QFTI.

The overall goal of the QFTD sign recovery circuit is to cast the ancilla qubit into a superposition of the form $\big(f_j\ket{0}_a +f'_j\ket{1}_a\big)\ket{j}$, where $f_j$ is the $j^\text{th}$ input function sample and $f'_j$ is the $j^\text{th}$ QFTD output.
Then, applying a Hadamard gate to the ancilla qubit results in the superposition state:
\begin{equation}\label{eq: Qftd_Signs_Goal}
    \bigg(\frac{f_j+f'_j}{\sqrt{2}}\ket{0}_a + \frac{f_j - f'_j}{\sqrt{2}}\ket{1}_a\bigg)\ket{j}
\end{equation}
The probabilities of observing the $\ket{0}_a\ket{j}$ or $\ket{1}_a\ket{j}$ states are given by $(f_j+f'_j)^2/2$ and $(f_j - f'_j)^2/2$, respectively.
Making use of the known inputs and the relative probabilities of the ancilla states, the signs of $f'_j$ may be obtained according to the following truth table.
\begin{table}[h!]
\center
\begin{tabular}{c c||c}
    \textbf{If} & \textbf{And} & \textbf{Then}\\ 
    \hline
    \multirow{2}{3em}{$f_j>0$} 
    & $(f_j+f_j')^2 > (f_j-f_j')^2$ & $f'_j>0$\\
    & $(f_j+f_j')^2 < (f_j-f_j')^2$ & $f'_j<0$\\
    \hline
    \multirow{2}{3em}{$f_k<0$} 
    & $(f_j+f_j')^2 > (f_j-f_j')^2$ & $f'_j<0$\\
    & $(f_j+f_j')^2 < (f_j-f_j')^2$ & $f'_j>0$\\
\end{tabular}
\caption{Truth table showing the different cases relating the input function's sign, the relative magnitudes of $\ket{0}_a\ket{j}$ and $\ket{1}_a\ket{j}$ states, and the output function's sign.}
\label{table: SignTable}
\end{table}

\newpage
In order to achieve the ancilla superposition given by Eq.\eqref{eq: Qftd_Signs_Goal}, the QFTD sign recovery circuit shown in Fig.~\ref{fig: Qftd_SignRecoveryCirc} is initialized with the same input as the original QFTD circuit.
An additional ancilla qubit, $\ket{b}$, is also initialized to the state $\ket{0}_b$.

By applying a Hadamard gate to the ancilla $b$ qubit and performing a controlled QFTD operation, the system is cast into the following state:
\begin{equation}
    \bigg[\frac{f_j}{\sqrt{2}}\ket{0}_b\ket{0}_a~+~\frac{1}{\sqrt{2}}\ket{1}_b\big(*\ket{0}_a + f'_j\ket{1}_a\big)\bigg]\ket{j}
\end{equation}
If the controlled QFTD operation does successfully encode the derivative, the ancilla qubit is shifted to the state $\ket{1}_a$.
This ancilla state proceeds to control a $CNOT$ gate on the $b$ qubit, resulting in the state:
\begin{equation}
    \ket{0}_b\bigg(\frac{f_j}{\sqrt{2}}\ket{0}_a + \frac{f'_j}{\sqrt{2}}\ket{1}_a\bigg)\ket{j} + *\ket{1}_b
\end{equation}
By applying an open-controlled Hadamard gate based on the ancilla $b$ , the system is cast into the following state:
\begin{equation}\label{eq: Qftd_Signs_Actual}
    \ket{0}_b\bigg[\frac{f_j+f_j'}{2}\ket{0}_a + \frac{f_j-f'_j}{2}\ket{1}_a\bigg]\ket{j} + *\ket{1}_b
\end{equation}
Finally, we see that $\ket{0}_b$ states in Eq.~\eqref{eq: Qftd_Signs_Actual} are in the form of Eq.~\eqref{eq: Qftd_Signs_Goal}.
At this time, the full quantum statevector is measured and $\ket{1}_b$ states are discarded while the $\ket{0}_b$ states are analyzed in accordance with Table~\ref{table: SignTable} to identify the sign of each QFTD output.
At the same time, the square-roots of the original QFTD outputs are post-processed to recover the correctly scaled results. 
By combining the results of each circuit after all post-processing steps, the full derivative, with proper scale and sign, is obtained.

As previously mentioned, this process may also be applied to extract signs from the QFTI output. In that case, the overall strategy remains the same; however, the QFTI's additional ancilla qubits must be treated appropriately. 
Noting that the QFTI is successful when the ancilla register is cast into the $\ket{001}_a$, the QFTI sign recovery circuit shown in Fig.~\ref{fig: QFTI_SignRecoveryCirc} prepares a superposition of the form $\ket{0}_b\big(f_j\ket{000}_a + I_j\ket{001}_a\big)\ket{j} + *\ket{1}_b$, where $I_j$ denotes the $j^\text{th}$ QFTI output.
Then, applying a controlled Hadamard gate to the least significant ancilla qubit creates the following superposition:
\begin{equation}\label{eq: QFTI_Signs_Actual}
    \ket{0}_b\bigg(\frac{f_j+I_j}{\sqrt{2}}\ket{000}_a + \frac{f_j - I_j}{\sqrt{2}}\ket{001}_a\bigg)\ket{j}+*\ket{1}_b
\end{equation}

Noting the similarities between the expressions Eq.~\eqref{eq: Qftd_Signs_Actual} and Eq.~\eqref{eq: QFTI_Signs_Actual}, the signs of the integration results may be obtained by comparing the amplitudes of the $\ket{0}_b\ket{000}_a$ and $\ket{0}_b\ket{001}_a$ states according to the same truth table as the QFTD, Table~\ref{table: SignTable}, but exchanging $f'_j$ with $I_j$.

\section{Numerical Examples}
This section presents a series of numerical examples that illustrates the performance of the QFTD and the QFTI algorithms on simple test functions.
The algorithms described above were implemented within the Qiskit environment \cite{javadi-abhari_quantum_2024}.
For all examples, functions are assumed to be unitless.
Performance and accuracy are evaluated by directly comparing circuit measurements with the square of the corresponding analytical solutions.
We emphasize that the primary results of the algorithms are not the measured quantities of $\big[\frac{df}{dx}\big]^2$ or $[I(x)]^2$; instead, the primary results of the QFTD and QFTI are their final state vectors whose basis state coefficients are correctly signed and directly proportional to the expected derivative or integral results.
In the following examples, the squared measurement results are provided only as a means of probing contents of the quantum statevector.

\subsection{QFTD Numerical Examples}\label{sec: QFTD_Examples}
To demonstrate the validity and performance of the QFTD algorithm, the function of interest was sampled at $N$ intervals of width $\Delta x$ throughout the domain. The samples were normalized by their $l^2$ norms, and Qiskit’s initialization function was used to encode the normalized samples into the state amplitudes of the $n$-qubit input register in Fig.~\ref{fig: QftdCircuitExample}. The number of qubits was set to $n=8$ and all numerical results were obtained using $10^7$ shots in Qiskit’s Aer Simulator \cite{javadi-abhari_quantum_2024}. Upon completion of the circuit simulations, observations with the ancilla qubit in state $\ket{1}_a$ were discarded. Binary states were converted to integers, denoted as $j$, and mapped to the correct $x$ value using the relationship $x_j = j\Delta x +\min(x)$. 
The probability amplitude of each state, $\psi ^2$, was obtained by dividing the state’s count by the total number of shots, and the correct amplitude of the QFT-based derivative was recovered using Eq.~\eqref{eq:  Qftd_AmplitudeRecovery}.

\begin{figure}[!h]
    \centering
    \includegraphics[width=0.75\textwidth]{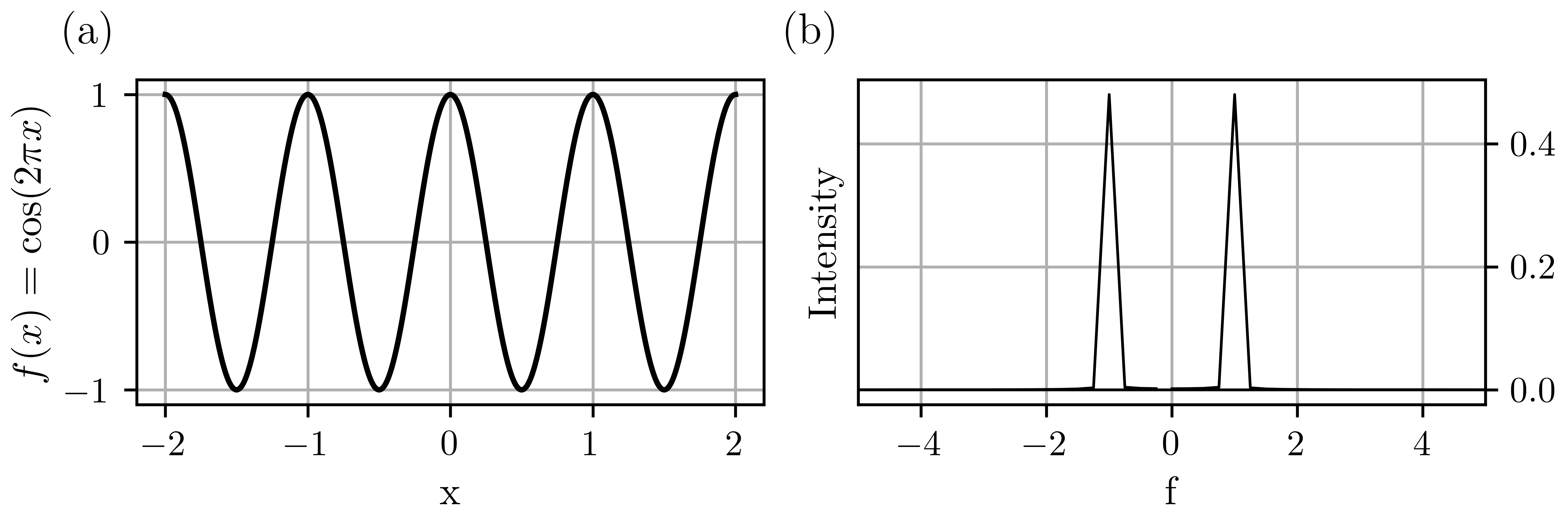}
    \caption{Illustration of (a) the input function $f(x) = \cos(2\pi x)$ and (b) its frequency spectrum.}
    \label{fig: Cos_FuncAndFreq}
\end{figure}

The first example presents the derivative of a trigonometric function $f(x) = \cos(2\pi x)$, where $x \in [-2, 2]$ and $\Delta x = \frac{2-(-2)}{2^8}\approx\num{1.6e-2}$. Both the function and its derivative are $C^{\infty}$ on $\mathbb{R}$. The function and its spectrum are presented in Fig.~\ref{fig: Cos_FuncAndFreq}.

\begin{figure}[!h]
    \centering
    \includegraphics[width=0.75\textwidth]{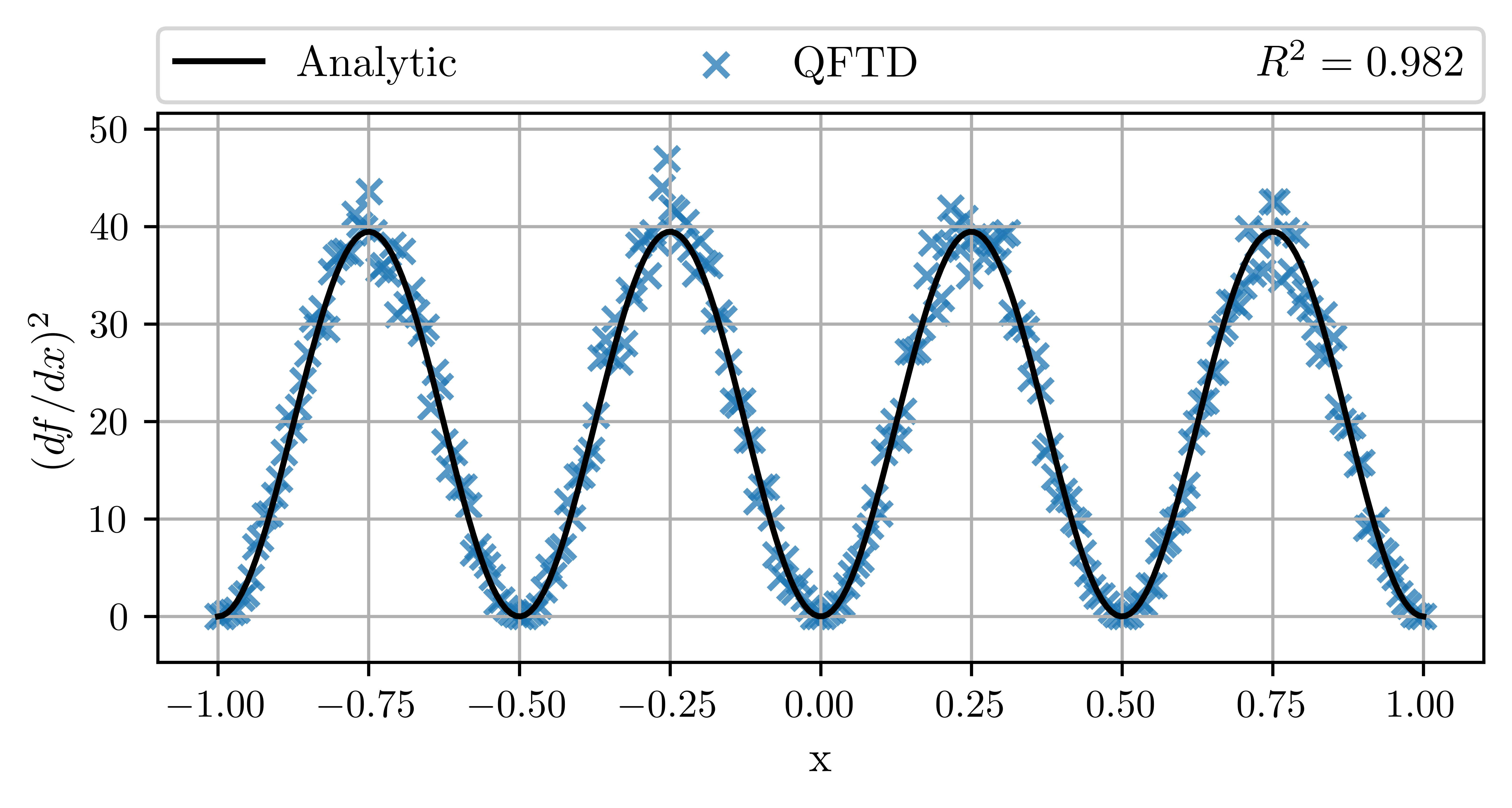}
    \caption{Squared amplitude of the first derivative of the input function $f(x)=\cos(2\pi x)$ obtained via the QFTD algorithm. The solid line provides the square of the analytical derivative for reference.}
    \label{fig: QFTD_cosPiX_Result}
\end{figure}

Figure~\ref{fig: QFTD_cosPiX_Result} shows a direct comparison between the analytical derivative and the one obtained via the QFTD quantum algorithm. Specifically, the output of the quantum circuit is compared to the square of the analytical derivative because the derivative is encoded in the quantum state's amplitude, and the square of the amplitude represents the state's measurement probability.
The graphical comparison suggests a good agreement between the analytical and the quantum-circuit-based results.
In more quantitative terms, this observation is also supported by the high value of the coefficient of determination $R^2=0.982$.
The quality of this result is aligned with the findings from classical DFT-based derivative algorithms \cite{johnson_notes_nodate, sunaina_calculating_2018}, which noted that periodic input functions are well-suited for Fourier-based derivative techniques.
Nonetheless, this result may be further improved by conducting circuit simulations with additional shots and by using additional qubits to increase the sample frequency of the input function.

\begin{figure}[!h]
    \centering
    \includegraphics[width=0.75\textwidth]{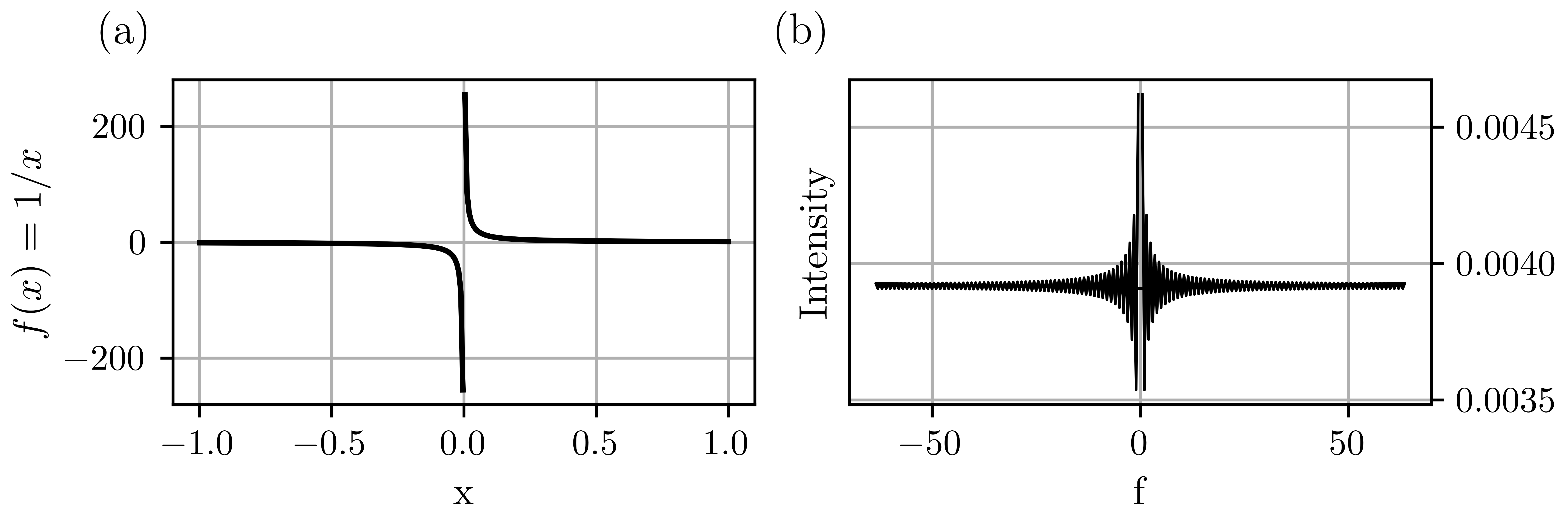}
    \caption{Illustration of (a) the input function $f(x) = 1/x$ and (b) its frequency spectrum. In the frequency plot, uniformly non-zero intensities at high frequencies indicate that DFTs and QFTs will truncate the high-frequency portions of the function's true frequency spectrum.}
    \label{fig: 1overX_FuncAndFreq}
\end{figure}

The second example is chosen to be $f(x) = 1/x$ (see Fig.~\ref{fig: 1overX_FuncAndFreq}). Also in this case, the function and its derivatives are $C^{\infty}$ over the restricted real axis $(-\infty,0)\cup(0,\infty)$. For the numerical example, we choose $x \in (-1,0^-)\cup(0^+,1)$ with $\Delta x = \frac{1-(-1)}{2^8}\approx\num{7.8e-3}$. This example is chosen to illustrate the ability of the QFTD algorithm to handle functions with singularities. In addition, the function and its derivatives vanish at infinity but are not compactly supported, hence determining a relatively broad spectral content. This latter aspect is important to acknowledge when using a Fourier transform-based approach because the number of harmonics retained in the transform is critical to properly capture the spectral content of the original signal.

\begin{figure}[!h]
    \centering
    \includegraphics[width=0.75\textwidth]{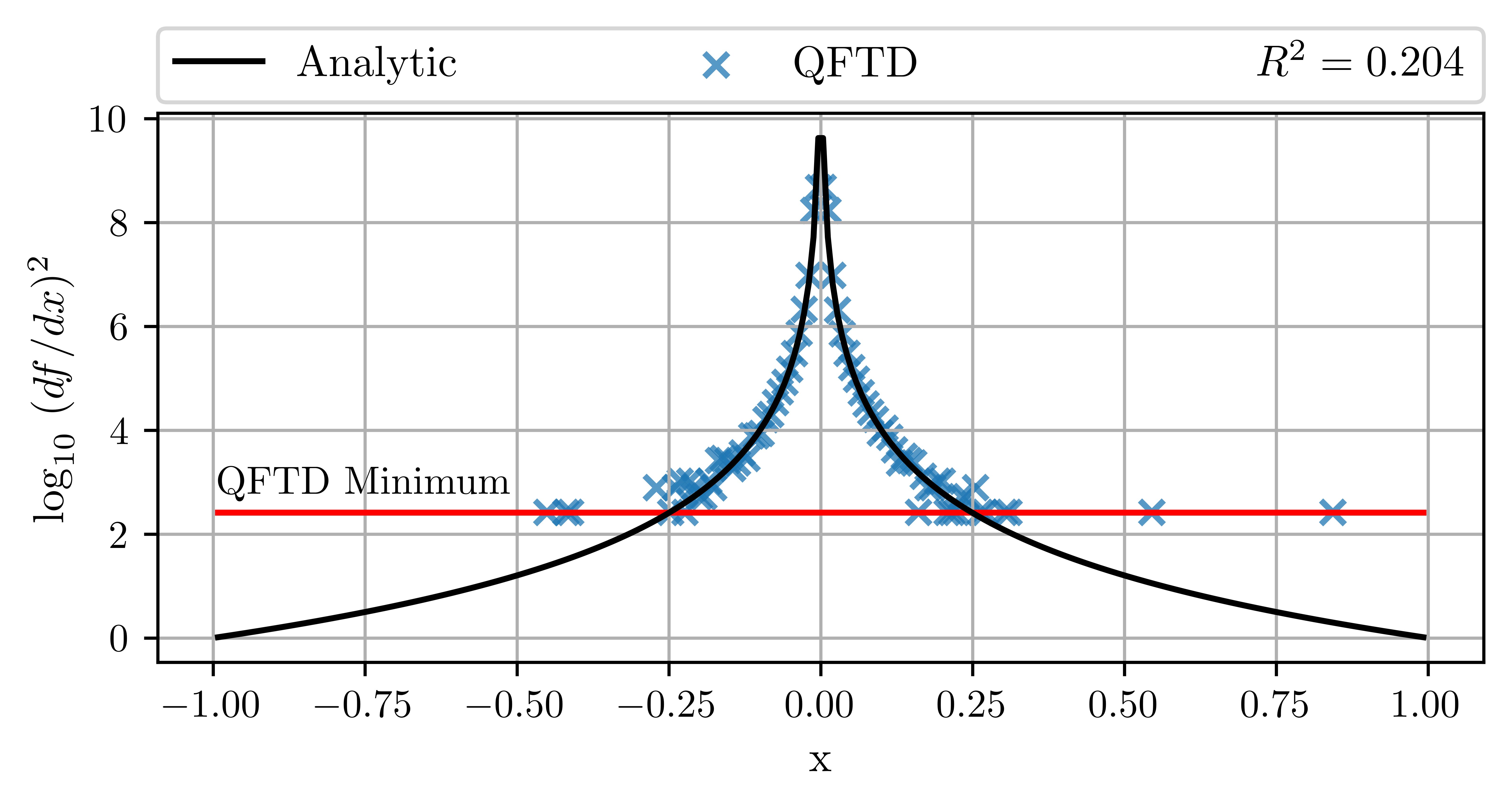}
    \caption{Squared amplitude of the first derivative of the input function $f(x)=1/x$ obtained via the QFTD algorithm. The solid line provides the square of the analytical derivative for reference.}
    \label{fig: QFTD_1OverX_Result_2sided_Full}
\end{figure}

\begin{figure}[!h]
    \centering
    \includegraphics[width=0.75\textwidth]{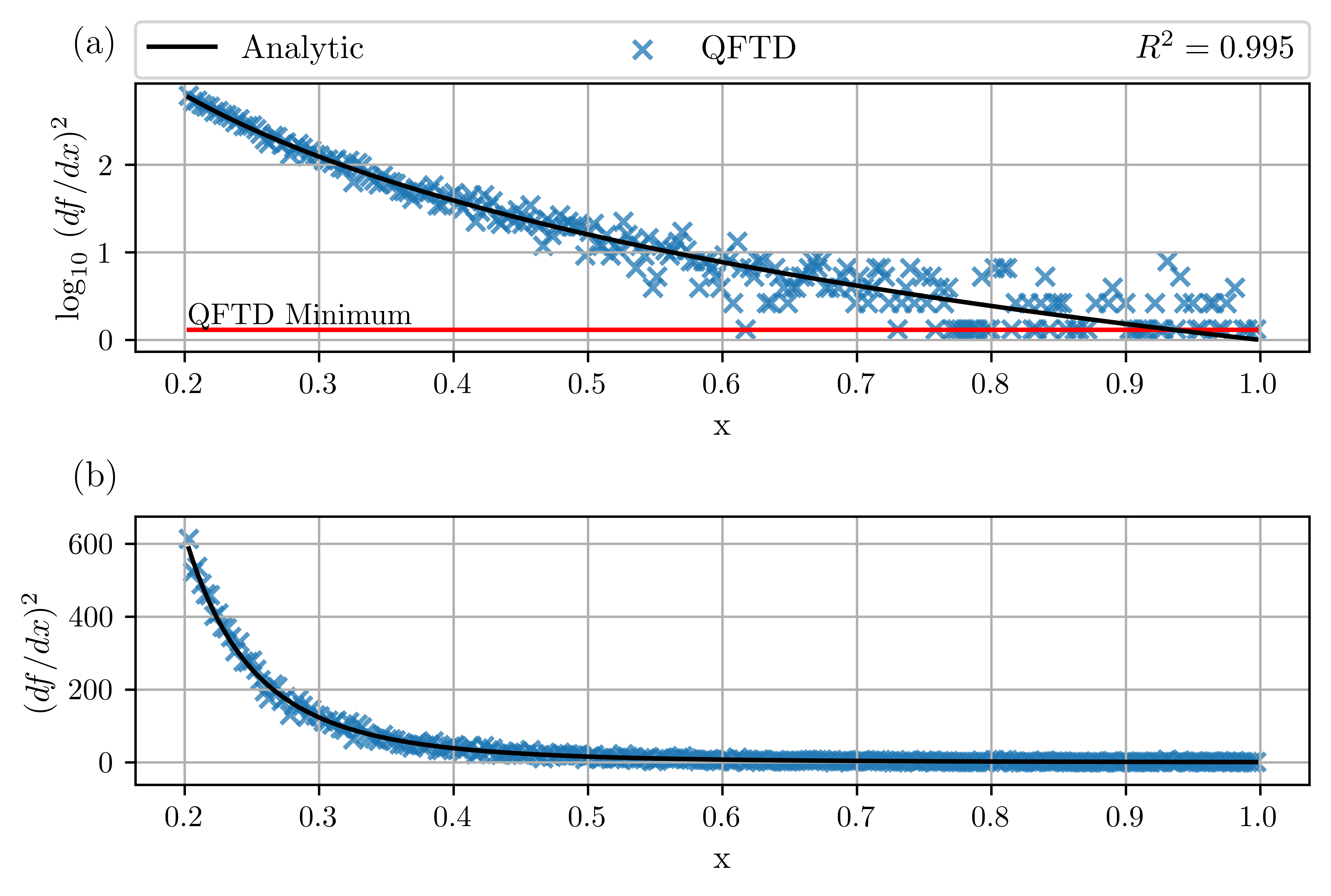}
    \caption{QFTD result for input function $f(x) = 1/x$ evaluated with $10^8$ shots over a shifted domain $[0.2, 1]$ compared with square of analytical derivative result presented with (a) semi-logarithmic scaling and (b) linear scaling.}
    \label{fig: QFTD_1OverX_Result_1sided_Full}
\end{figure}

Figure~\ref{fig: QFTD_1OverX_Result_2sided_Full} shows in semi-logarithmic scale the direct comparison of the QFTD and the corresponding analytical derivative for an input function $f(x)=-1/x^2$. As in the previous case, results are compared in terms of the square of the amplitude of the resulting function.
Results in Fig.~\ref{fig: QFTD_1OverX_Result_2sided_Full} show the impact of the total shot count, function scaling, and input domain on the accuracy of the QFTD algorithm.
For a simulation using $10^7$ shots, the minimum probability with which a state can be measured is $1/10^7$.
Using this example's values of $\Delta x$ and $||f||_{l^2}$ and scaling the minimum probability according to Eq.~\eqref{eq: Qftd_Resolution}, the QFTD resolution is found to be $r \approx 260$.
Indicated by a horizontal line in Fig~\ref{fig: QFTD_1OverX_Result_1sided_Full} the base-ten logarithm of the minimum resolution is found to be $2.4$.
In regions where the analytical derivative is less than the QFTD's minimum resolution, the quantum result is assigned a value of zero, and the results are not included in Fig.~\ref{fig: QFTD_1OverX_Result_2sided_Full}.
Resulting in the low correlation between quantum and analytical results indicated by $R^2 = 0.204$, only $25\%$ of the sampled points are expected to be observed using this example's QFTD resolution.
 
In Fig.~\ref{fig: QFTD_1OverX_Result_1sided_Full}, we report the calculation for the same function $f(x) = 1/x$ over the restricted computational domain $x\in[0.2, 1]$, with a total of $10^8$ shots. By shifting the domain, we maintain the asymptotic behavior near zero; but, evidenced by the $l^2$ norm reduction from $||f||_{l^2}\approx400$ to $||f||_{l^2}\approx36$, the scale of the input is reduced. Additionally, the restricted domain decreases the value of $\Delta x$ from $\Delta x \approx 7.8e^{-3}$ to $\Delta x \approx 3.1e^{-3}$ and the increased shot count reduces the minimum observable probability by a factor of $10$. With these parameters, the QFTD resolution indicated by Eq.~\eqref{eq: Qftd_Resolution} is $r \approx 1.3$ and approximately $92\%$ of the sample points are expected to be captured in the QFTD output.

Noting the agreement between the quantum and the analytical results indicated by the $R^2$ value of $0.995$, we conclude that the QFTD algorithm accurately estimates the derivative of the input function. As expected, the graph shows asymptotic behavior close to the lower bound of the computational domain while the rest of the function is now more accurately estimated because shots are not all clustered around the high amplitude region close to the singularity. While the semi-log plot of Fig.~\ref{fig: QFTD_1OverX_Result_1sided_Full}(a) contains striations near the domain's upper bound that result from the algorithm's resolution, visual inspection of the linearly scaled results in Fig.~\ref{fig: QFTD_1OverX_Result_1sided_Full}(b) show the apparent inaccuracy is exacerbated by logarithmic scaling. 

\begin{figure}[!h]
    \centering
    \includegraphics[width=0.75\textwidth]{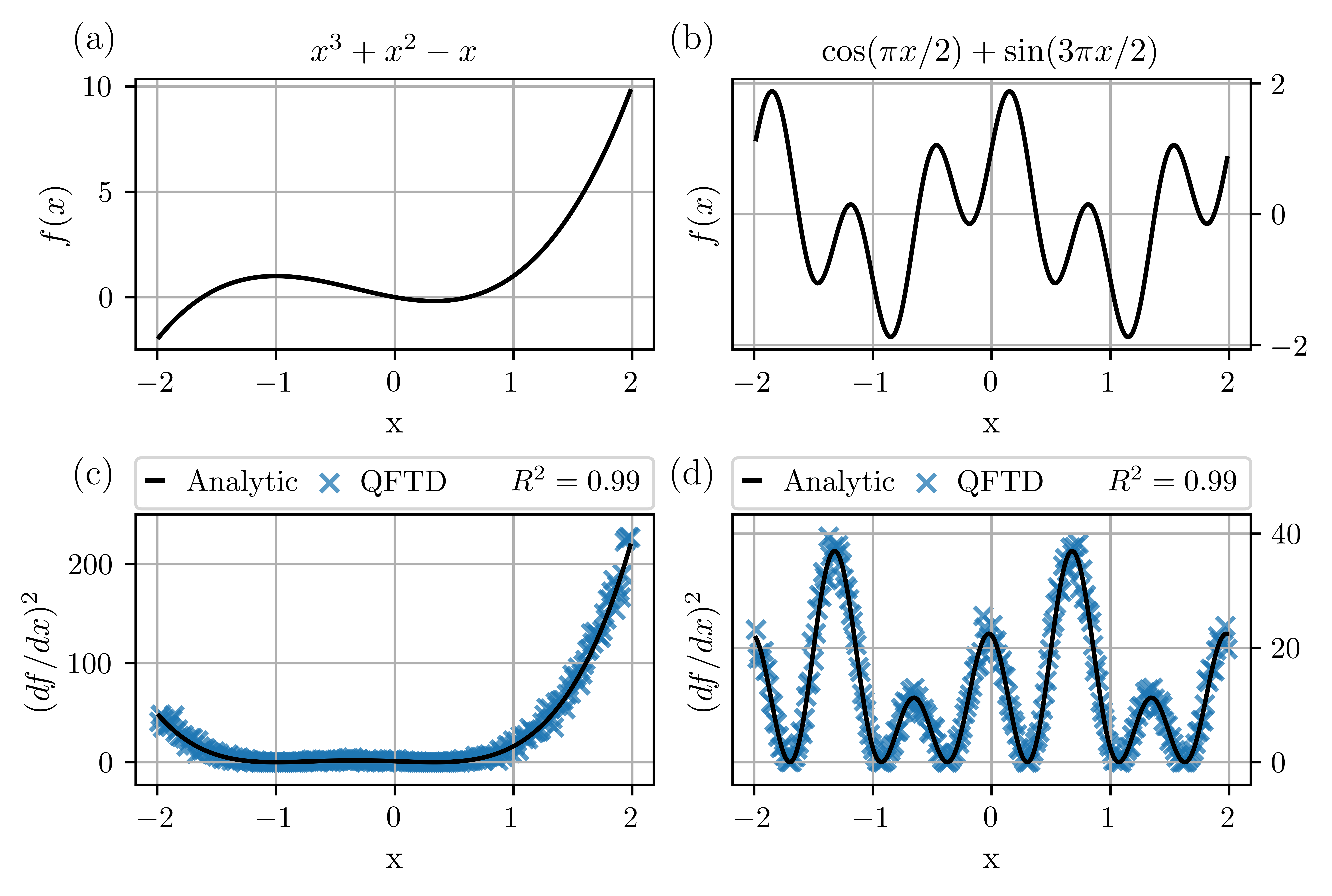}
    \caption{Results of the derivative algorithm applied to a polynomial input function (a) $f(x) = x^3 +x^2 -x$ and (b) a trigonometric input function with multiple harmonics $f(x) = \cos(\frac{\pi x}{2})+\sin(\frac{3\pi x}{2})$. The comparison between the QFTD and the analytical derivatives is presented in (c) for the polynomial function and in (d) for the trigonometric function.}
    \label{fig: QFTD_AdditionalExamples}
\end{figure}

To further demonstrate the robustness of the QFTD algorithm, two additional examples are briefly presented. Figure~\ref{fig: QFTD_AdditionalExamples} shows the performance of the QFTD algorithm when applied to a polynomial function, $f(x)=x^3 +x^2 +x$, and a trigonometric function containing two harmonics, $f(x)=\cos(\frac{\pi x}{2}) + \sin(\frac{3\pi x}{2})$. The QFTD results for both functions are in close agreement with theoretical expectations, as also indicated by the respective $R^2$ values of $0.99$ and $0.99$. 

\subsection{QFTI Numerical Examples}\label{sec: QFTI_Examples}
The numerical assessment of the QFTI algorithm was performed on the same test functions introduced above.
For the larger QFTI circuits, Qiskit's Statevector Sampler\cite{javadi-abhari_quantum_2024} was used to analyze circuits because its usage of random access memory enabled faster circuit simulations.
The implementation procedure for the QFTI algorithm is nearly identical to that used for QFTD, with two primary exceptions. 
First, based on the bit ordering illustrated in Fig.~\ref{fig: QftiCircuit}, successful circuit runs occur when the first three qubits are in the state $\ket{001}$. 
Therefore, upon measuring the circuit, only these states are kept for further analysis.
Second, because two additional qubits are required to perform the cumulative summations, the number of qubits used to store function samples is restricted to 6 for simulation efficiency. Thus, QFTI simulations presented herein employ 6 qubits in the primary register and 3 qubits in the ancilla register.

\begin{figure}[!h]
    \centering
    \includegraphics[width=0.75\textwidth]{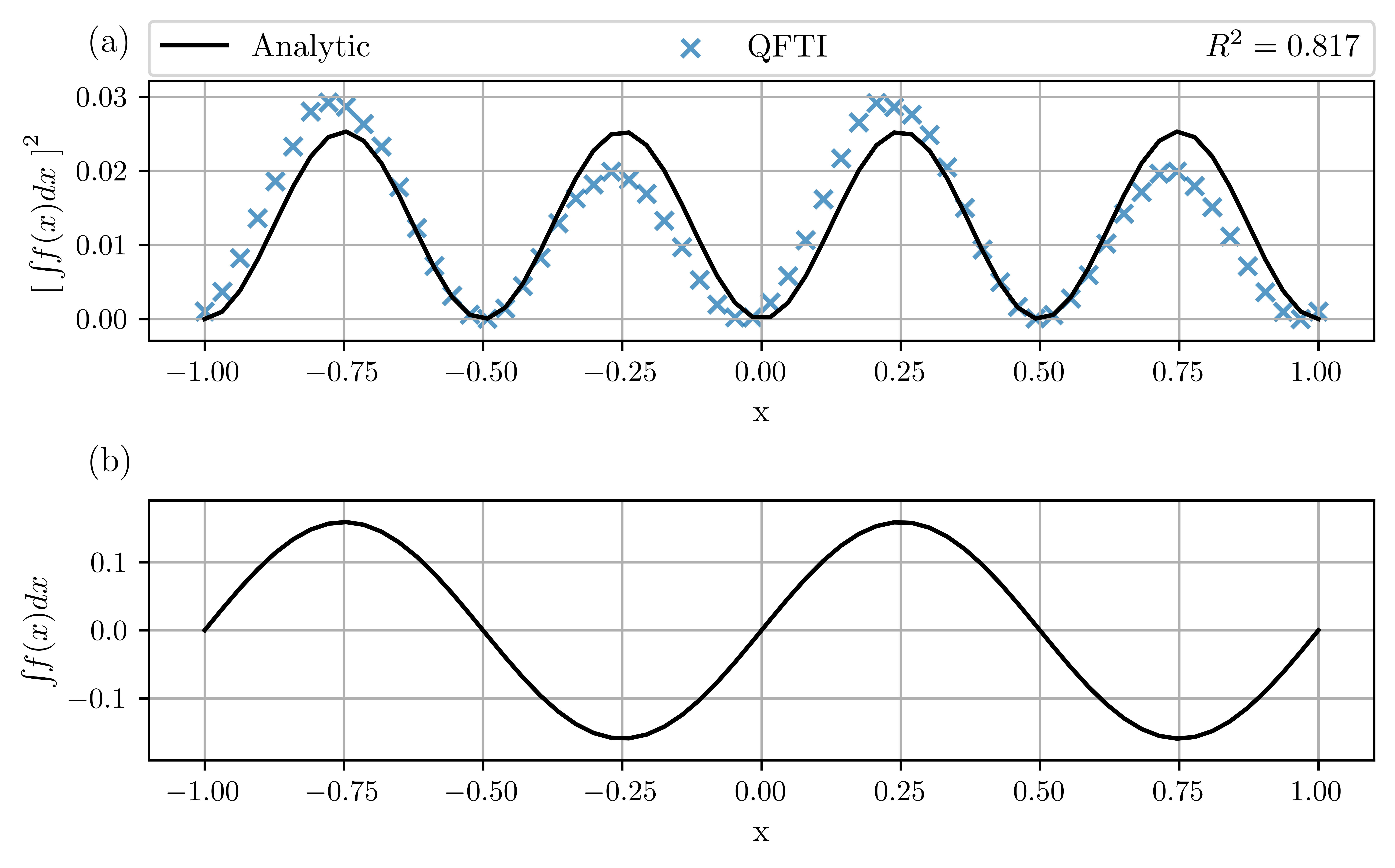}
    \caption{(a) The QFTI result for the input function $f(x) = \cos(2\pi x)$ is compared with the analytical result for the partially bound integral $\int_{x_0}^x \cos(2\pi x) dx$. (b) The analytical integral is displayed without squaring the amplitude to illustrate the respective correlations between QFTI over/underestimation and positivity/negativity of the analytical integral.}
    \label{fig: QFTI_cosPiX_Result}
\end{figure}

Figure~\ref{fig: QFTI_cosPiX_Result}(a) presents the QFTI result for the function $f(x) = \cos(2\pi x)$. As previously done, results are compared against the square of the analytical result $\int_{x_0}^x \cos(2\pi x) dx=\frac{1}{2\pi}\sin(2\pi x) - \frac{1}{2\pi}\sin(2\pi x_0)$.
The true integral is expected to coincide with the line $y=0$ at $x_0$ when including the integration constant given by $-\frac{1}{2\pi}\sin(2\pi x_0)$; however, the quantum algorithm exhibits a systemic error caused by the initial point.
When the input function evaluated at $x_0$ is non-zero, the corresponding differential area encoded by the QFTI algorithm is also non-zero.
When the cumulative summations are performed, the erroneous initial area propagates throughout the result, ultimately causing uniform overestimates or underestimates. 
Figure~\ref{fig: QFTI_cosPiX_Result}(b) presents the analytical result without squaring to illustrate the circumstances inducing over or underestimation. Noting where the analytical result is positive or negative, it is apparent that overestimation corresponds to regions where the analytical integral is positive. In regions where the analytical integral is negative, the QFTI's overestimation results in a smaller absolute value which, upon squaring, yields a smaller final result. 

\begin{figure}[!h]
    \centering\includegraphics[width=0.75\textwidth]{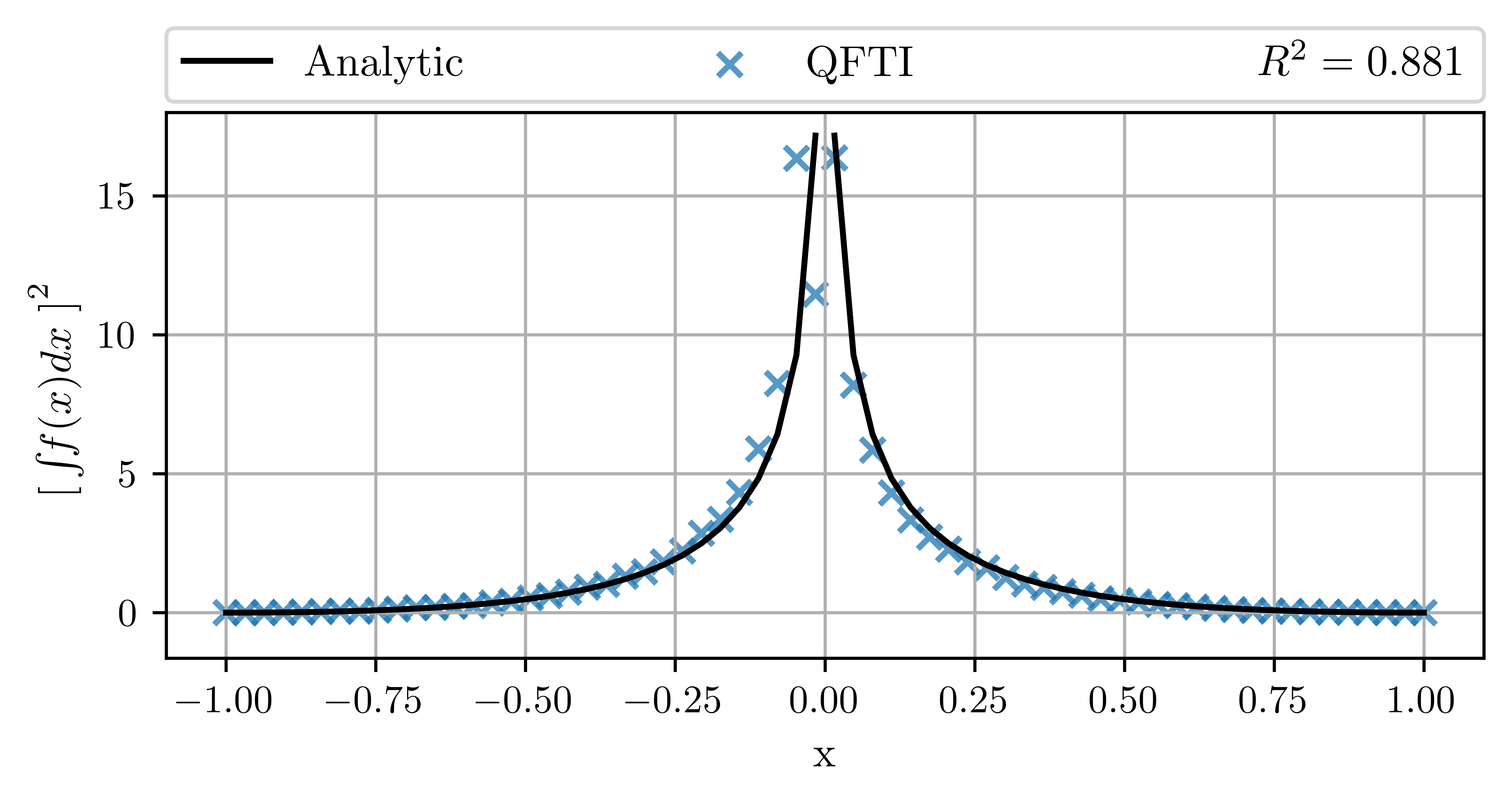}
    \caption{The QFTI result for the input function $f(x) = 1/x$ is compared with the analytical result for the partially bound integral $\int_{x_0}^x 1/x~dx$}
    \label{fig: QFTI_1OverX_2sided_Result}
\end{figure}

In Fig.~\ref{fig: QFTI_1OverX_2sided_Result}, the QFTI results are presented for the input function $f(x) = 1/x$. In this case, over and underestimation of the expected value results from an erroneous negative initial area. Because the initial error is negative, the final result demonstrates underestimation in regions where the integral is positive, while overestimation occurs where the analytical integral is negative. Regardless of the initial error's sign, its influence is directly related to the step size. As $\Delta x$ approaches zero, the first differential area approaches zero causing a smaller offset between the numerical and analytical integrals.

\begin{figure}[!h]
    \centering
    \includegraphics[width=0.75\textwidth]{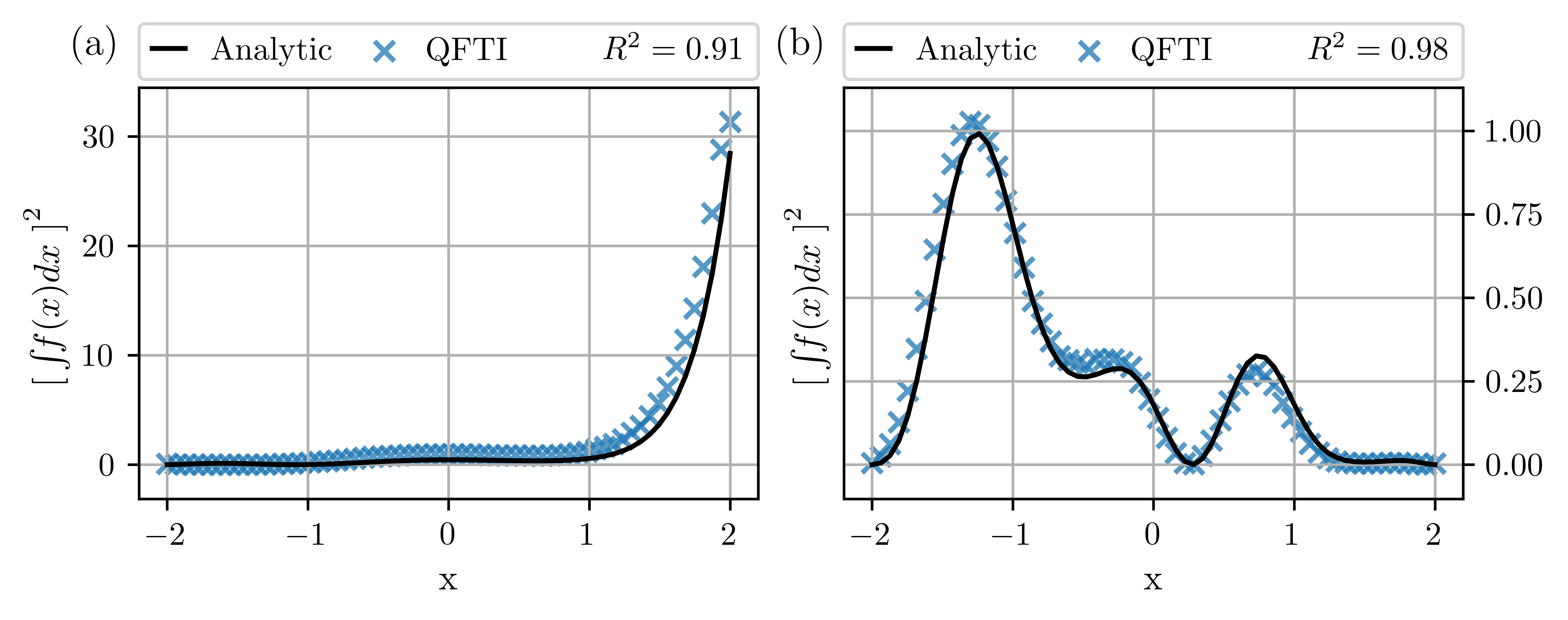}
    \caption{Six-qubit examples comparing the QFTI algorithm's output against the square of the analytically expected integral for (a) the input function $f(x) = x^3+x^2-x$ and (b) the input function $f(x) = \cos(\pi x/2)+\sin(3\pi x/2)$.}
    \label{fig: QFTI_AdditionalInputs}
\end{figure}

Again, two additional examples are briefly provided to illustrate the versatility of the QFTI algorithm. 
Using the same input functions presented in Fig.~\ref{fig: QFTD_AdditionalExamples}(a) and Fig.~\ref{fig: QFTD_AdditionalExamples}(b), the QFTI algorithm is implemented to estimate the integrals of a polynomial function and a multi-harmonic trigonometric function. Comparing the two results in Fig.~\ref{fig: QFTI_AdditionalInputs}, both examples agree with theoretical expectations as quantified by the $R^2$ values of $0.91$ and $0.98$ for the polynomial and trigonometric functions, respectively. 

\subsection{QFTD-Based Gradient Estimation}\label{sec: ndQFTD_Examples}
In the following example, a two-dimensional implementation of the QFTD algorithm is applied to estimate the magnitude of a two-dimensional function's gradient. 
The scalar function $f(y,x) = \cos(\pi x)\sin(\pi y/2)$ is discretized in the domain $x,y \in[-2,2]$ by $n^{(x)}=6$ qubits along the $x$ direction and $n^{(y)}=6$ qubits along the $y$ direction to form a grid of $N_x\times N_y$ input function samples, where $N_x=N_y=2^6$ and $\Delta x=\Delta y = \frac{2-(-2)}{2^6}=\num{6.25e-2}$. 
The grid of function samples is reshaped and encoded as a normalized vector in accordance with Eq.~\eqref{eq: ndEncoding}. For reference, the encoded state takes the form:
\begin{equation}
    \ket{\Psi_0}=\frac{1}{||f||_{l^2}}
        \begin{pmatrix}
        f\big(y_0,~x_0\big)\\
        f\big(y_0,~x_1\big)\\
        \vdots \\
        f\big(y_1,~x_0\big)\\
        f\big(y_1,~x_1\big)\\
        \vdots \\
        f\big(y_{N_y-1},~x_{N_x-2}\big)\\
        f\big(y_{N_y-1},~x_{N_x-1}\big)\\
    \end{pmatrix}
\end{equation}

Qiskit's Aer Simulator is employed to simulate the $6\times6$ qubit equivalent of the two-dimensional QFTD circuit shown in Fig.~\ref{fig: ThreeQubit_NdQFTD_Circ} using $M=10^8$ shots. 
Upon measuring the statevector, partial derivative results are identified according to Table~\ref{table: MixedPartials}.
Like the one-dimensional case, the measured outputs, $\psi_k^2$, are rescaled according to Eq.~\eqref{eq: Qftd_AmplitudeRecovery}.
Finally, the magnitude of the gradient is classically calculated by evaluating the expression $|\nabla f(y,x)|=\sqrt{\big(\frac{\partial f}{\partial x}\big)^2+\big(\frac{\partial f}{\partial y}\big)^2}$.

As was the case for the one-dimensional QFTD, the measured result for each partial derivative is proportional to the square of the theoretical partial derivative.
Therefore, for the sake of investigating algorithm accuracy, only the gradient magnitudes are considered.
Comparing the theoretical and quantum results in Fig.~\ref{fig: ndQftd}, a high $R^2$ value of $0.975$ is observed.
Similar to the one-dimensional QFTD results, this indicates that the multidimensional QFTD algorithm functions as expected, yielding outputs proportional to the theoretical derivative. 
Although this example only considers gradient magnitudes for the sake of investigating algorithm accuracy, the gradient's absolute value also conveys important information in the form of critical points, where the gradient's absolute value is $0$.
Furthermore, because the correct sign information is available prior to measurement, this example illustrates that the QFTD algorithm may effectively serve as a subroutine in future quantum algorithms that require gradient estimations as input.

\begin{figure}[!h]
    \centering
    \includegraphics[width=1\textwidth]{Images/NdQftd_cosPiX_x_sinPion2Y_N6M6.png}
    \caption{Comparison of (a) the $M=10^8$ shot, $6\text{-qubit}\times6\text{-qubit}$ QFTD gradient result, and (b) the theoretical gradient magnitude of the input function $f(x,y) = \cos(\pi x)\sin(\pi y /2)$.}
    \label{fig: ndQftd}
\end{figure}

\begin{figure}[!h]
    \centering
    \includegraphics[width=0.5\linewidth]{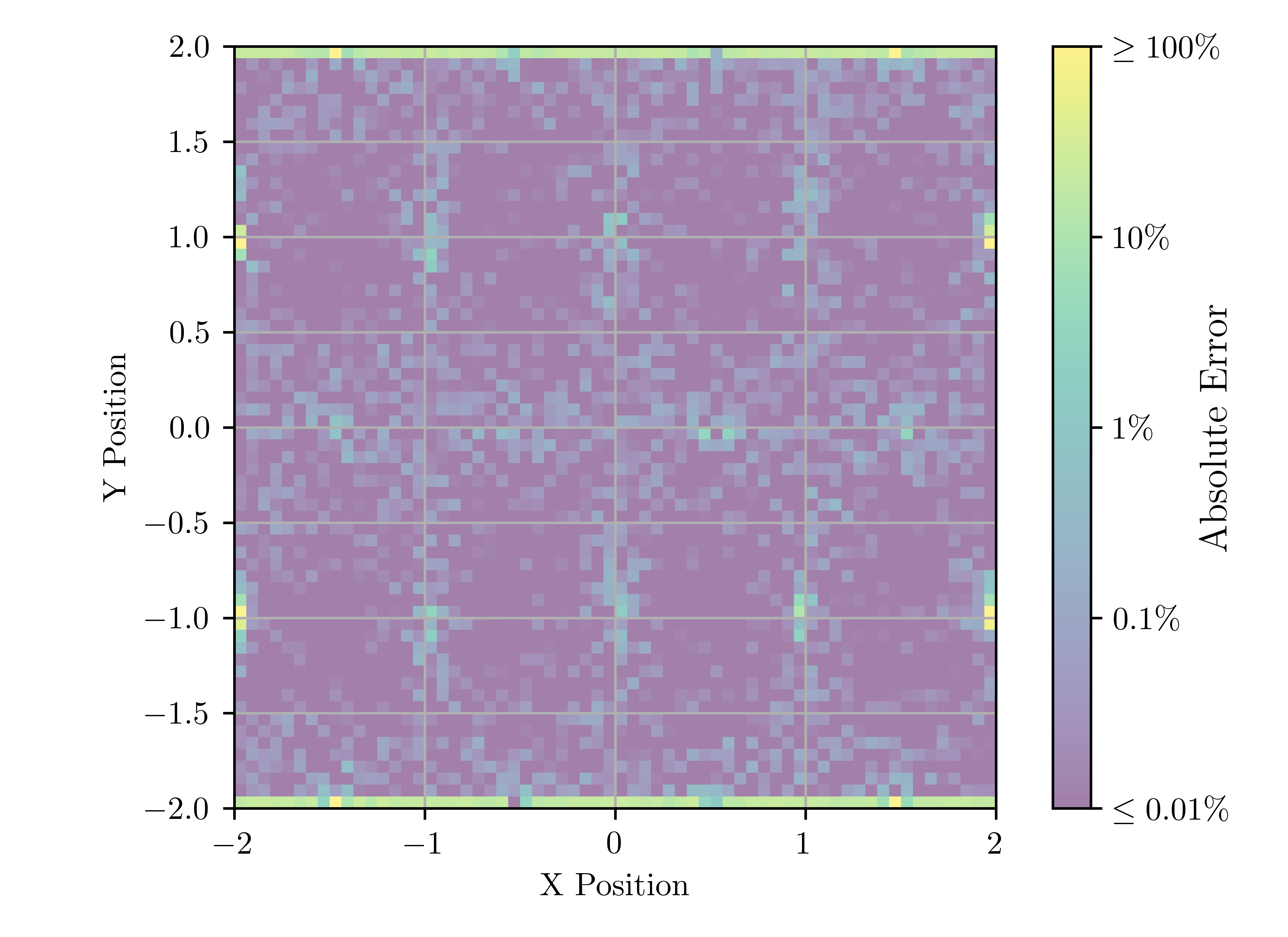}
    \caption{Logarithmically scaled error map comparing the actual gradient magnitude and the QFTD gradient magnitude estimate for the input function $f(x,y) = \cos(\pi x)\sin(\pi y /2)$.}
    \label{fig: ndQFTD_ErrorMap}
\end{figure}

To further illustrate the accuracy of the two-dimensional QFTD,  Fig.~\ref{fig: ndQFTD_ErrorMap} presents a logarithmically scaled error map comparing the QFTD output with the actual gradient magnitude.
Complementing the large $R^2$ value indicated by Fig.~\ref{fig: ndQftd}, the error map suggests strong agreement between the theoretical and quantum results.
However, this figure also reveals significant errors near domain boundaries.
Such errors are the result of Gibbs phenomena, and are expected when implementing spectral techniques.
These errors may be mitigated using standard spectral analysis techniques such as windowing.
Comparing Fig.~\ref{fig: ndQFTD_ErrorMap} and Fig.~\ref{fig: ndQftd}, it is also observed that small errors appear near the minima of the gradient magnitude.
These errors occur as a result of truncation when probabilities fall below the QFTD resolution.
Recalling Eq.~\eqref{eq: Qftd_Resolution}, these truncation errors may be mitigated by increasing the number of shots used in circuit simulations.

\subsection{Sign Recovery Examples}
In this section, we incorporate the sign extraction procedure to simulate the two final examples provided in \S~\ref{sec: QFTD_Examples} and \S~\ref{sec: QFTI_Examples} for the QFTD and QFTI algorithms.
Simulation parameters such as the shot count, input domain, and sample size are kept consistent with the original results.

\begin{figure}[!h]
    \centering
    \includegraphics[width=0.75\textwidth]{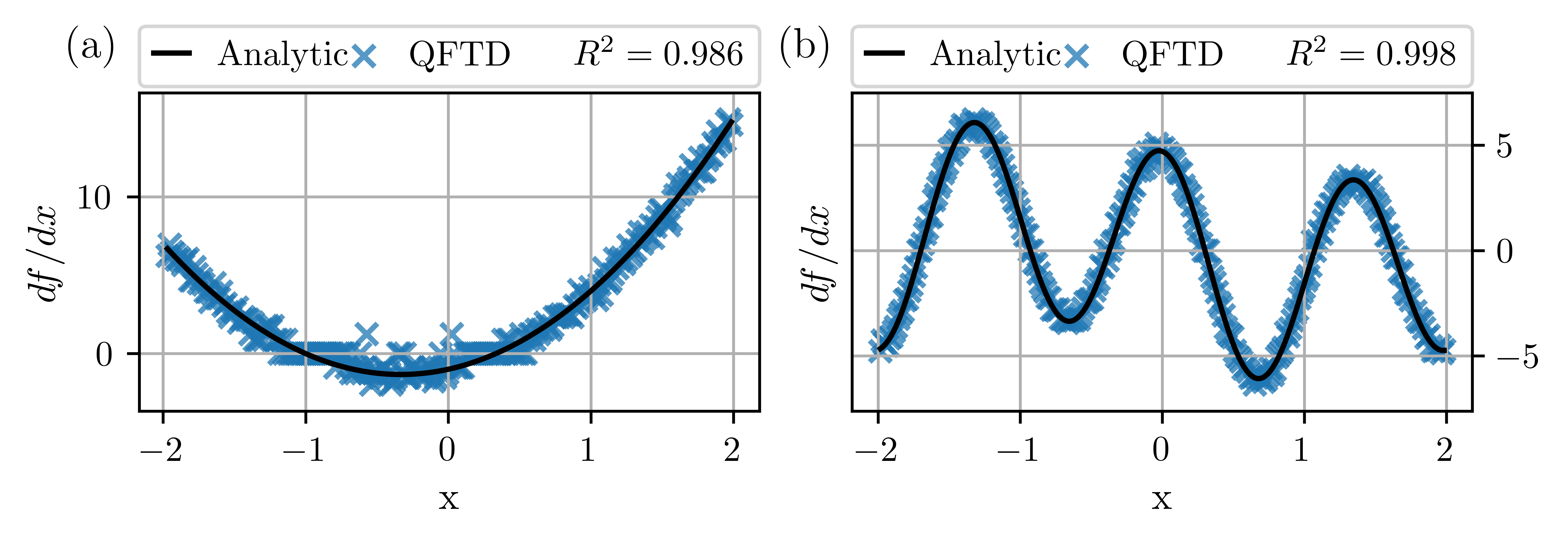}
    \caption{Eight-qubit examples (excluding ancilla qubits) comparing the QFTD algorithm's signed output against the analytically expected derivative for (a) the input function $f(x) = x^3+x^2-x$ and (b) the input function $f(x) = \cos(\pi x/2)+\sin(3\pi x/2)$.}
    \label{fig: SignedExamples_QFTD}
\end{figure}
\begin{figure}[!h]
    \centering
    \includegraphics[width=0.75\textwidth]{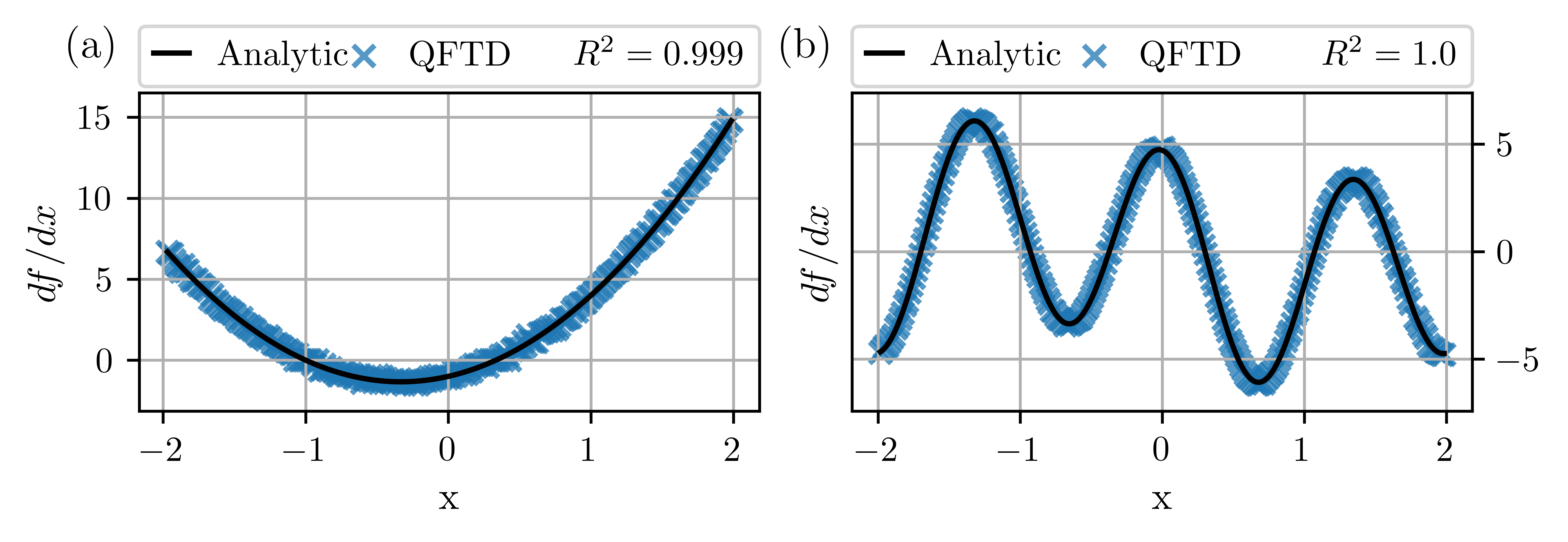}
    \caption{Eight-qubit examples comparing the QFTD algorithm's signed output against the analytically expected derivative with resolution improvements obtained by increasing simulation shots from $M=10^7$ to $M=10^8$.}
    \label{fig: SignedExamples_QFTD_Ns10e8}
\end{figure}

Figure~\ref{fig: SignedExamples_QFTD} shows the QFTD results with sign recovery applied for (a) $f(x) = x^3+x^2-x$ and (b) $f(x) = \cos(\pi x/2)+\sin(3\pi x/2)$. In both cases, the quantum results closely follow the expected trend of the analytical predictions. 
In Fig.~\ref{fig: SignedExamples_QFTD}(a), regions near $x=-1$ and $x=0$ show that the QFTD result is truncated to 0, while this behavior was not observed in the unsigned results.
This occurs because the resolution of the QFTD algorithm is insufficient to distinguish between the different ancilla cases outlined in Table~\ref{table: SignTable} when evaluating signs.
As shown in Fig.~\ref{fig: SignedExamples_QFTD_Ns10e8}, these resolution induced errors are eliminated by increasing the total number of shots from $M=10^7$ to $M=10^8$.

Noting that the signs of the quantum results agree with expectations in Fig.~\ref{fig: SignedExamples_QFTD} and Fig.~\ref{fig: SignedExamples_QFTD_Ns10e8}, these examples provide additional confirmation that the QFTD algorithm encodes a state vector with coefficients directly proportional to the expected numerical derivative.

\begin{figure}[!h]
    \centering
    \includegraphics[width=0.75\textwidth]{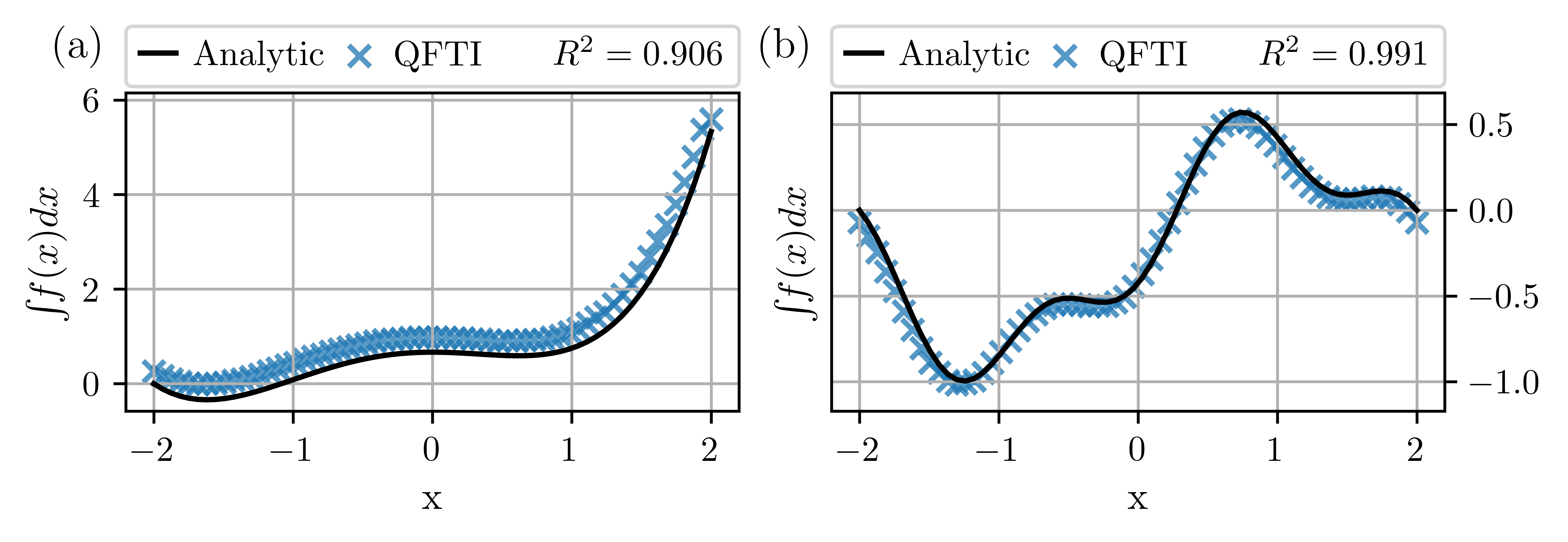}
    \caption{Six-qubit examples comparing the QFTI algorithm's signed output against the analytically expected integral for (a) the input function $f(x) = x^3+x^2-x$ and (b) the input function $f(x) = \cos(\pi x/2)+\sin(3\pi x/2)$.}
    \label{fig: SignedExamples_QFTI}
\end{figure}

Figures~\ref{fig: SignedExamples_QFTI}(a),(b) show the QFTI results with sign recovery applied for $f(x) = x^3+x^2-x$ and $f(x) = \cos(\pi x/2)+\sin(3\pi x/2)$, respectively. Noting that the recovered sign consistently agrees with analytical predictions, these plots further demonstrate that the QFTI procedure encodes a correctly signed quantum state vector that is proportional to the expected integral.

The constant vertical offset visible between the QFTI output and the expected result further illustrates the error that is accumulated due to an incorrect initial area.
Recall from previous QFTI examples that this erroneous initial area propagates throughout the QFTI results.
Because previous results only considered the squared output, these errors caused either overestimation or underestimation of the actual integral, depending on the sign of the initial error and the actual integral at a given point.
Given that Fig.~\ref{fig: SignedExamples_QFTI} shows the actual integral values, the erroneous initial area estimate only contributes to a constant vertical offset.
As noted previously, this error is linearly dependent on $\Delta x$; therefore, the error may be controlled by sampling the input function with a step size.

\section{Discussion}

The previous numerical results illustrated the validity and effectiveness of the QFTD and QFTI algorithms for a variety of different test functions. In the following, we provide additional discussions of each algorithm's error scaling, complexity, and overall differences with respect to other methods.

\subsection{Error Analysis} 
While assessing the practical potential of the quantum numerical calculus techniques presented in this work, the error associated with each algorithm, $\epsilon$, is also of interest.
To draw generalizable conclusions regarding the accuracy of each algorithm, we consider implementing the algorithms using 1) an ideal quantum computer, and 2) a shot-based, fault-tolerant quantum computer.

\begin{figure}[!h]
    \centering
    \includegraphics[width=0.99\textwidth]{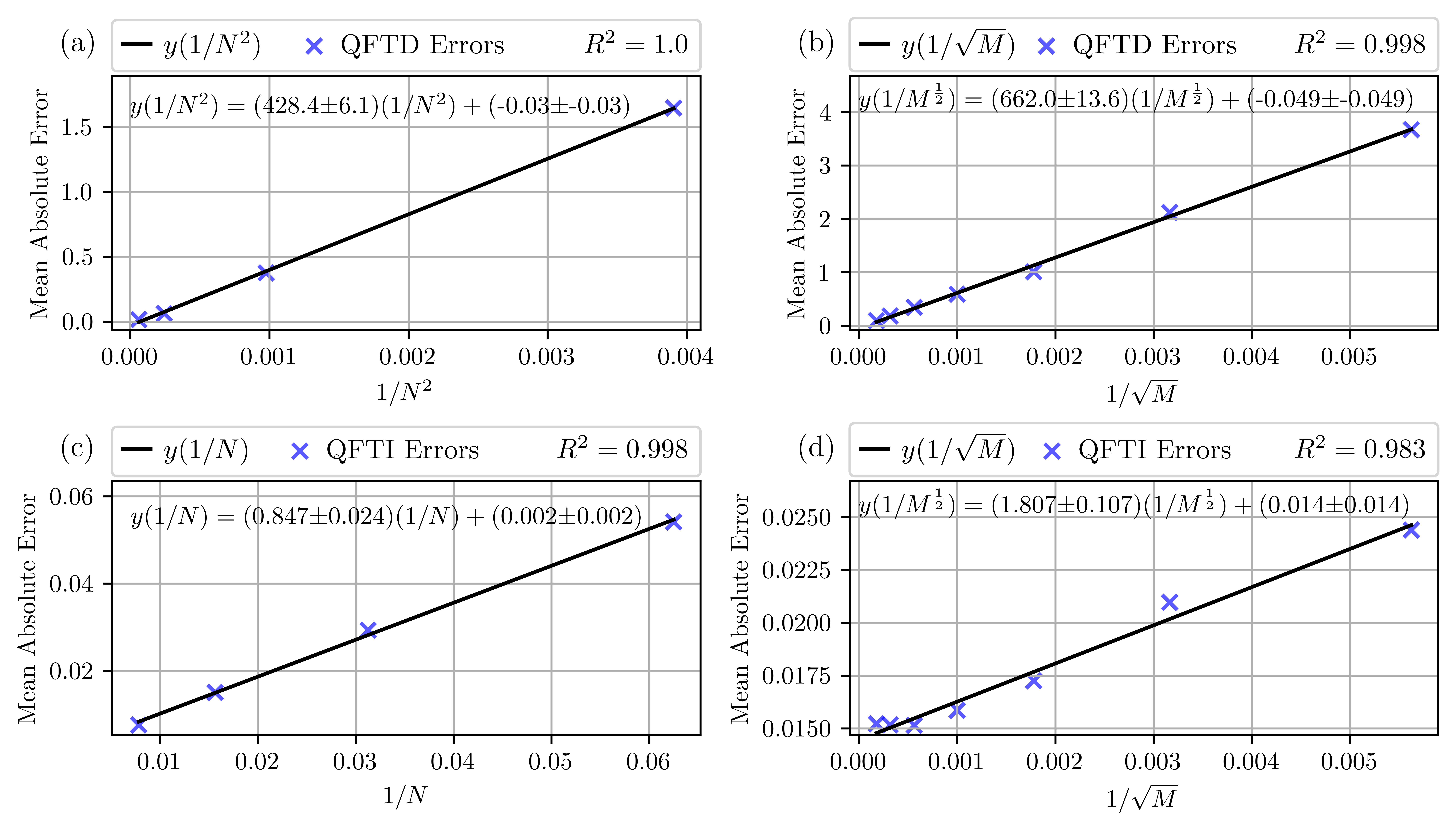}
    \caption{Trend of the mean absolute error observed for (a),(b) the QFTD algorithm and (c),(d) the QFTI algorithm. Linear regressions reveal that sample size dependent errors are given by the lines $y(1/N^2)$ or $y(1/N)$ for the QFTD and QFTI, respectively. Both algorithms have shot-dependent errors given by the line $y(1/\sqrt{M})$.}
    \label{fig: ErrorTrends}
\end{figure}

Assuming access to an idealized quantum computer, the state amplitudes may be perfectly known.
This is equivalent to performing traditional shot-based computations with an infinite number of shots.
Under this assumption, we recall Eq.~\eqref{eq: CentralDiff} and Eq.~\eqref{eq: TrapMeth}, which indicate that the QFTD and QFTI algorithms are quantum implementations of the central difference differentiation and trapezoidal integration schemes, respectively.
Thus, each derivative estimate returned by the QFTD algorithm is expected to possess an error of $\epsilon=\mathcal{O}\big(\Delta x^2\big)$ or, equivalently, $\epsilon=\mathcal{O}\big(N^{-2}\big)$ \cite{chapra_numerical_2005}.
Similarly, the QFTI algorithm is expected to possess an error of $\mathcal{O}\big(\Delta x^2 N\big)$ or, equivalently, $\mathcal{O}\big(N^{-1}\big)$ \cite{strang_36_2016}. Indeed, the respective errors of order $\mathcal{O}\big(N^{-2}\big)$ and $\mathcal{O}\big(N^{-1}\big)$ for the QFTD and QFTI are observed in Fig.~\ref{fig: ErrorTrends} which depicts linearized graphs of the mean absolute errors from circuit simulations with varying qubit amounts.

In the realistic case of shot-based quantum computations, the accuracy of state amplitude measurements is affected by the total number of shots employed.
Additionally, the number of shots required to capture a function's behavior depends strongly on the function itself.
For instance, a function with significant magnitude variations such as $f(x) = 1/x$, where $f\rightarrow\infty$ as $x\rightarrow0$ and $f(x)\rightarrow 0$ as $x\rightarrow\infty$, requires a large number of shots to simultaneously observe high-probability and low-probability behavior.
Because the shot counts required to accurately represent a result is problem dependent, we consider the shot-based nature of quantum computers as a source of error. 
Then, the shot-dependent algorithm resolutions given by Eq.~\eqref{eq: Qftd_Resolution} and Eq.~\eqref{eq: QFTI_Resolution} may be treated as variance estimates.
It follows that the standard error of each algorithm is proportional to the resolution's square root, $\epsilon\propto \sqrt{r}$. 
For the QFTD algorithm, Eq.~\eqref{eq: Qftd_Resolution} suggests that shot-dependent errors may be estimated as $\epsilon=\mathcal{O}(M^{-1/2})$.
Similarly, in the case of the QFTI algorithm, Eq.~\eqref{eq: QFTI_Resolution} suggests that shot-dependent error may be estimated as $\epsilon=\mathcal{O}(M^{-1/2})$.

Accounting for both the central difference and the shot-dependent errors, the QFTD error may be written as:
\begin{equation}\label{eq: QFTD_Error}
    \epsilon = \max\bigg(\mathcal{O}\big(N^{-2}\big),\qquad \mathcal{O}\big(M^{-1/2}\big)\bigg)
\end{equation}
Similarly, when both error sources are considered for the QFTI algorithm, the error is found to be:
\begin{equation}\label{eq: QFTI_Error}
    \epsilon = \max\bigg(\mathcal{O}\big(N^{-1}\big),\qquad \mathcal{O}\big(M^{-1/2}\big)\bigg)
\end{equation}

In Fig.~\ref{fig: ErrorTrends}, the error scalings indicated by Eq.~\eqref{eq: QFTD_Error} and Eq.~\eqref{eq: QFTI_Error} are depicted by linearized graphs of the mean absolute error.
Error data is obtained by applying the QFTD to the input function $f(x) = \cos(2\pi x)$ for $x\in[-2, 2]$, while QFTI error data is obtained using the same input function for $x\in[-2, 2]$.
In Fig.~\ref{fig: ErrorTrends}(a), the QFTD algorithm's sample size dependent error is depicted by the errors of a $10^8$-shot QFTD simulation with varying sample sizes.
Similarly, Fig.~\ref{fig: ErrorTrends}(c) illustrates the QFTI's sample size dependent error using results from a $10^7$-shot QFTI simulation with varying sample sizes.
In Fig.~\ref{fig: ErrorTrends}(b), the QFTD algorithm's shot-dependent error is illustrated by the errors of a $10^8$-qubit QFTD simulation with varying shot counts.
Similarly, Fig.~\ref{fig: ErrorTrends}(d) illustrates the QFTI's sample size dependent error using results from a $10^6$-qubit QFTI simulation with varying shot counts.
Linear regressions reveal that sample size dependent errors are given by the lines $y(1/N^2)$ or $y(1/N)$ for the QFTD and QFTI respectively, while both algorithms have shot-dependent errors are given by the line $y(1/\sqrt{M})$.

\subsection{Complexity Analysis}
\noindent Using two applications of the QFT circuit with gate complexity $\mathcal{O}\big(\log N\big)$, and $n = \log_2 N$ applications of $R_x$ gates, the gate-level complexity of the QFTD algorithm is $\mathcal{O}\big(\log N\big)$.
Because the complexity of algorithmic differentiation is problem dependent and varies with the quantum implementations of arithmetic operations\cite{henrard_algorithmic_2017,colucci_quantum_2021}, we do not draw conclusions regarding its efficiency relative to the QFTD algorithm.
The VQE approach carries a query complexity of $\mathcal{O}(n^4N_\theta/\epsilon^2)$ for an $n$-qubit circuit with $N_\theta$ parameters\cite{mitarai_theory_2020}, and Jordan's algorithm estimates $d$-dimensional gradients with query complexity $\mathcal{O}(\sqrt d/\epsilon)$\cite{jordan_fast_2005}. 
Noting that these algorithms necessarily employ oracles to implement expressions derived from the input functions, only their query complexities are precisely known while their gate-level complexities must also include the cost of oracle implementations.
To further contrast against the previous quantum differentiation algorithms, the QFTD's complexity depends only on the number of samples, $N$, and not the target error, $\epsilon$.
Because the complexities of these algorithms each depend on fundamentally different quantities, direct comparisons of algorithmic complexity are insufficient to describe their relative efficiencies.

In the case of the quantum integration algorithm, the complexity of the QFT operations and the wavenumber rotations are identical to the QFTD's $\mathcal{O}(\log N)$ complexity. However, the partial summation operation must also be taken into account.
The additional unitary operation adds a single layer to the circuit’s overall depth.
This gives the QFTI a query complexity of $\mathcal{O}(\log N)$; however, the gate level complexity introduced by the partial summation gate $U_H$ is not clearly defined. Indeed, only a lower bound of $\mathcal{O}\big(\log N\big)$ may be used to estimate the QFTI algorithm's gate-level complexity.
Compared with the latest QMCI technique's query complexity of $\mathcal{O}(\sqrt{N})$\cite{shu_general_2024}, the QFTI algorithm presented in this paper is exponentially more efficient than the latest QMCI technique.
Unlike the QFTI or QMCI, the quantum Riemann sum method's gate-level complexity \textit{is} known to be $\mathcal{O}\big(\log N\big)$ \cite{shu_general_2024}.
Therefore, it is evident that the quantum integration algorithm presented in this work is not always the most efficient technique for quantum numerical integration of function samples.

\newpage

\subsection{Highlights of the Proposed Method}

To facilitate the comparison between different algorithms, Table~\ref{table: DerivativeCompTable} and Table~\ref{table: IntegralCompTable} summarize query complexities and required prior knowledge for each differentiation and integration approach.

\begin{table}[h!]
\begin{center}
\begin{tabular}{|c|c c c|}
    \hline
    \textbf{Technique} & \textbf{Prior Knowledge}& \textbf{Complexity} & \textbf{Output Type}\\ 
    \hline
    \textbf{AD}\cite{colucci_quantum_2021} & Algebraic Function & Problem Dependent & Point Estimate\\
    \hline
    \textbf{VQE}\cite{mitarai_theory_2020} & Algebraic Function & $\mathcal{O}(n^4 N_\theta/\epsilon^2)$ & Point Estimate\\
    \hline
    \textbf{Jordan's Algorithm}\cite{jordan_fast_2005} & Algebraic Function & $\mathcal{O}(\sqrt{d} / \epsilon)$  & Point Estimate\\
    \hline
    \textbf{QFTD} & Function Samples & $\mathcal{O}(\log N)$ & Domain-Wide Estimate\\
    \hline
    \rowcolor{lightgray}
    \textbf{Central Difference} & Function Samples  & $\mathcal{O}(N)$ & Domain-Wide Estimate\\
    \hline
\end{tabular}
\caption{Summary of input and output types, and complexities of available quantum differentiation techniques and, indicated with gray background, classical central difference differentiation.
Algorithmic differentiation (AD) is not ascribed a complexity because the operation decomposes arbitrary functions into unique combinations of elementary arithmetic operations whose quantum implementation contributes to the overall complexity. The VQE complexity assumes a parameterized circuit of $n$ qubits, with $N_\theta$ parameters and a target accuracy of $\epsilon$, while the complexity of Jordan's algorithm depends only on the target accuracy of $\epsilon$. For the QFTD and central difference methods, respective complexities are logarithmically and linearly dependent on $N$, the total number of function samples.}
\label{table: DerivativeCompTable}
\end{center}
\end{table}

\begin{table}[h!]
\center
\begin{tabular}{|c|c c c|}
    \hline
    \textbf{Technique} & \textbf{Prior Knowledge} & \textbf{Complexity} & \textbf{Output}\\ 
    \hline
    \textbf{QMCI}\cite{shu_general_2024} & Algebraic Function & $\mathcal{O}(\sqrt{N})$ &   Point Estimate\\
    \hline
    \textbf{Riemann Summation}\cite{shukla_efficient_2024} & Function Samples & $\mathcal{O}(\log N )$ & Point Estimate\\
    \hline
    \textbf{QFTI} & Function Samples & $\mathcal{O}(\log N )$  & Domain-Wide Estimate\\
    \hline
    \rowcolor{lightgray}
    \textbf{Trapezoidal Method} & Function Samples & $\mathcal{O}(N)$ & Domain-Wide Estimate\\
    \hline
\end{tabular}
\caption{Summary of inputs, outputs, and complexities of available quantum integration techniques and, shown in gray, classical trapezoidal integration. For the integration techniques, all complexities depend only on the number of input function samples, or, in the case of QMCI, the number of domain discretizations employed during the numerical integration process.}
\label{table: IntegralCompTable}
\end{table}

To the author's best knowledge, there are currently no other quantum differentiation algorithms that, like the QFTD, are capable of processing input samples to provide domain-wide derivative estimates.
As highlighted by Table~\ref{table: DerivativeCompTable}, we find that the QFTD operates in a fundamentally different manner than the AD, VQE approaches, or Jordan's algorithm, using function samples rather than function oracles, and returns domain-wide results rather than derivative or gradient estimations at a single point.
Because of their different functionalities, we find that complexity analysis is the only relevant argument when describing the advantage of the QFTD algorithm over quantum derivative algorithms.
Turning to classical algorithms, when the QFTD is compared with techniques offering similar functionality, such as the central difference method (i.e. the QFTD's classical analogue), Table~\ref{table: DerivativeCompTable} reveals that the quantum derivative algorithm provides an exponential efficiency improvement.

Turning to the QFTI algorithm, Table~\ref{table: IntegralCompTable} suggests that the three quantum differentiation algorithms previously mentioned also have different use cases.
The QMCI technique is useful when the function to be integrated is known a priori and the desired result has the form of a definite integral over a specified domain.
In contrast, the QFTI is useful when the input is only available numerically (that is via numerical samples, as in the case of experimental or numerical data) and the desired result has the form of a semi-definite integral with a fixed lower boundary (i.e. $\int_{x_0}^x f dx = F(x) - C, \text{where } C = F(x_0)$).
Finally, in cases where the input is only known via function samples and only a definite integral between two points is required, the Riemann sum technique is generally more efficient.
Analogously to the QFTD, when the QFTI is compared with its classical counterpart (i.e. trapezoidal integration), Table~\ref{table: IntegralCompTable} shows the QFTI and Riemann summation both introduce an exponential efficiency improvement; however, to estimate integral values at all $N$ points, replicating the result of the QFTI, the Riemann summation approach must be applied $N$ times, resulting in a gate complexity of $\mathcal{O}(N\log N)$.

\section{Conclusions}
We presented a quantum implementation of a DFT-based algorithm that enables either numerical differentiation or integration of a generic function. Starting from a uniform sampling of the function, the algorithm returns derivative estimates at discrete points selected within the domain of interest.
For an input consisting of $N$ function samples, the quantum derivative algorithm was found to have a complexity of $\mathcal{O}\big(\log N\big)$ and an error of $\max\big[ \mathcal{O} \big( N^{-2} \big), \quad \mathcal{O} \big( M^{-1/2} \big) \big]$.
Higher dimensional extensions of this technique were also presented. Interestingly, they are able to simultaneously evaluate all partial derivatives of a given function without affecting gate complexity.
Expanding on the spectral technique presented for function differentiation, a quantum algorithm for partially bound trapezoidal integration was also developed.
The integration method was found to have a minimal complexity of $\mathcal{O}\big(\log N\big)$, and an error of $\max\big[ \mathcal{O} \big( N^{-1} \big), \quad \mathcal{O} \big( M^{-1/2} \big) \big]$.

Up to normalization and scaling by known constants, the QFTD and QFTI algorithms both successfully encode the derivatives and integrals in the quantum state vectors.
Because sign information is lost during the measuring process, we further substantiated the efficacy of the QFTD and QFTI algorithms by presenting a post-processing technique that enables the explicit recovery of sign information.
We note that this post-processing procedure is only necessary when the user requires explicit access to the sign. Indeed, even without applying this technique, the signed results are still properly encoded in the quantum state vector.
This encoding is one of the primary results of this work and lays the foundation for the QFTD and QFTI algorithms to serve as core subroutines in applied quantum computing operations such as image processing, data analysis, and machine learning.

\section{Data Availability Statement}
Data and analysis supporting the findings of this article were generated by original codes which have been made publicly available \cite{cioni_qftd_2026}.

\newpage
\bibliography{QftdPaper_Refs}

\end{document}